\documentclass[a4paper,12pt]{article}
\usepackage{amsfonts,amsthm,amsmath,amssymb,graphicx,hyperref,youngtab}
\newcommand{\cH}{\mathcal H}

\newcommand{\cM}{\mathcal M}
\newcommand{\cN}{\mathcal N}
\newcommand{\cO}{\mathcal O}

\newcommand{\be}{\begin{equation}}
\newcommand{\bea}{\begin{eqnarray}}
\newcommand{\ee}{\end{equation}}
\newcommand{\eea}{\end{eqnarray}}

\begin{document}

\makeatletter
\@addtoreset{equation}{section}
\makeatother
\renewcommand{\theequation}{\thesection.\arabic{equation}}

\rightline{QMUL-PH-12-08}
\rightline{WITS-CTP-092}
\vspace{1.8truecm}

\vspace{15pt}

%%%%%%%%%%%%%%%%%

{\LARGE{  % \bf  Permutations, Strings and Feynman Graphs  
\centerline{   \bf A double coset ansatz } 
\centerline {\bf  for integrability in AdS/CFT } 
}}  

\vskip.5cm 

\thispagestyle{empty} \centerline{
    {\large \bf Robert de Mello Koch
${}^{a,} $\footnote{ {\tt robert@neo.phys.wits.ac.za}}}
   {\large \bf and Sanjaye Ramgoolam
               ${}^{b,}$\footnote{ {\tt s.ramgoolam@qmul.ac.uk}}   }
                                                       }

\vspace{.4cm}
\centerline{{\it ${}^a$ National Institute for Theoretical Physics ,}}
\centerline{{\it Department of Physics and Centre for Theoretical Physics }}
\centerline{{\it University of Witwatersrand, Wits, 2050, } }
\centerline{{\it South Africa } }

\vspace{.4cm}
\centerline{{\it ${}^b$ Centre for Research in String Theory, Department of Physics},}
\centerline{{ \it Queen Mary University of London},} \centerline{{\it
    Mile End Road, London E1 4NS, UK}}

\vspace{1.4truecm}

%%%%%%%%%%%%%%%%%
\thispagestyle{empty}

\centerline{\bf ABSTRACT}

\vskip.4cm 

We give a proof that the expected counting of strings attached 
to giant graviton branes in $AdS_5 \times S^5$, 
as constrained by the Gauss Law, matches the dimension spanned by 
the expected dual operators in the gauge theory.
The counting of string-brane configurations 
is formulated as a graph counting problem, which can be expressed 
as the number of points on a double coset involving permutation 
groups. Fourier transformation on the double coset suggests 
an ansatz for the diagonalization of the one-loop dilatation 
operator in this sector of strings attached to giant graviton 
branes. The ansatz agrees with  and extends recent results
which have found  the dynamics of open string excitations of giants  to be given by harmonic oscillators.  We prove 
that it provides the conjectured  diagonalization
leading to harmonic oscillators.

\setcounter{page}{0}
\setcounter{tocdepth}{2}

\newpage

\tableofcontents

\setcounter{footnote}{0}

\linespread{1.1}
\parskip 4pt

{}~
{}~

\section{Introduction}

The AdS/CFT correspondence \cite{malda} 
gives an equivalence between  $\cN =4$ super-Yang-Mills (SYM) 
theory in four dimensions and ten dimensional string theory 
in $AdS_5 \times S^5$. 
This allows the  construction of  quantum states in the  $\cN=4$ 
super-Yang-Mills \cite{bbns,cjr}, which are dual to half-BPS rotating branes 
(giant gravitons \cite{mst,myers,hash}) 
 in the string theory. The construction of states uses the 
representation theory of symmetric and unitary groups, and in particular 
inter-relations between them encoded in Schur-Weyl duality.  
States $\chi_R ( Z ) $ are associated with Young diagrams $R$. When 
$R$ has  order one rows of length order $N$, they 
are dual to  multiple giants consisting of large branes 
in the $AdS_5$ space. States associated with long columns are dual 
to branes large in the $S^5$ space. The construction of 
states corresponding to strings attached to giants was undertaken  
in \cite{Balasubramanian:2002sa,Aharony:2002nd,Berenstein:2003ah,Sadri:2003mx,
Berenstein:2005fa,Berenstein:2006qk,Balasubramanian:2004nb,dssi,dssii,bds}.  
A particularly simple limit arises when the lengths of the columns 
are separated by 
order $N$ and the action of the one-loop dilatation operator simplifies 
significantly \cite{Koch:2011jk,
Koch:2010gp,gs,DeComarmond:2010ie,Carlson:2011hy,Koch:2011hb}. 
The diagonalization of the one-loop dilatation operator 
reveals a new integrable sector, with the appearance of 
harmonic oscillator spectra describing excitations 
of strings attached to the giants. In this new sector
both planar and non-planar diagrams contribute at
large $N$. Integrability in the planar limit was discovered in\cite{mz,bks}
and is discussed in the recent review\cite{intreview}.

The half-BPS giants are constructed from 
products of traces of powers of one matrix $Z$.
General multi-trace operators made from $n$ copies of $Z$
 can be parameterized by permutations, since 
upper indices are some permutation of the lower indices. 
The Young diagram operators are obtained by summing 
over permutations, with weights given by characters 
in irreducible representations (irreps) $R$ of the permutations. 
If there are $n$ copies of $Z$ 
involved, then the permutations are in the symmetric group 
$S_{n}$ of order $n!$. Permutations related by conjugation 
give the same trace, so conjugacy classes of  $S_n$ give a
natural parameterization of traces. Going to a representation basis 
of Young diagrams gives a simple way to implement finite $N$ relations, 
by restricting the  Young diagrams to have no more
than $N$ rows \cite{cjr}.

The states for attached strings can be constructed by replacing 
some of the $Z$  matrices with another ``impurity'' matrix $Y$. Each matrix 
$Y$ generates a 1-bit string \cite{Giles:1977mpa,Berenstein:2002jq,Balasubramanian:2002sa,Vaman:2002ka} 
with angular momentum in the $y$-direction. 
If there are $n$ copies of $Z$ involved and $m$ of $Y$, 
then the traces are parameterized by permutations $S_{m+n}$, 
but there are equivalences under conjugation by elements 
in $S_n \times S_m$.  Representation theory again gives a 
natural basis $\chi_{R,(r,s)\mu\nu } ( Z ,  Y ) $
 for these conjugacy classes, which naturally 
incorporates finite $N$ effects. The labels include 
$R$, a Young diagram corresponding to an irrep of 
 $S_{m+n}$ and a pair $(r,s)$ 
of Young diagrams for  $S_n \times S_m$. There are additional 
multiplicity labels $\mu , \nu$, which each run over the multiplicity 
with which $(r,s)$ appears when the irrep $R$ of $S_{m+n}$ is 
decomposed under the action of the subgroup $S_{n} \times S_m$. 
 
Recent progress in the study of perturbations 
$\chi_{R,(r,s)\mu \nu } ( Z ,  Y ) $  of 
giants $\chi_R ( Z ) $
has found that the calculation of the spectrum of the 
one-loop dilatation operator 
reduces to systems of harmonic oscillators. The harmonic oscillator dynamics 
consists of motion of $p$ particles along the real line, 
their coordinates being given by the lengths of the Young diagram $R$ (which has $p$ long rows or long columns) interacting via quadratic 
pair-wise interaction potentials. Arriving at this harmonic 
oscillator dynamics requires a diagonalization in the space of 
labels $ ( s , \mu , \nu )$. There are $U(1)^p$ conserved charges in the 
system which forces the Young diagram $r$ to be completely determined by  $R$.

This diagonalization in the $ ( s , \mu ,\nu ) $ sector has been achieved 
in various special cases in earlier papers. The numerical studies of  
\cite{Koch:2010gp,DeComarmond:2010ie} considered $m=2,3,4$ $Y$s and demonstrated a linear spectrum. 
An analytic approach which solved the problem when $R$ has 2 rows or columns and $m$ is general 
was given in \cite{Carlson:2011hy} for operators built from 2 scalars $Z,Y$
and in \cite{Koch:2011jk} for operators built using 3 scalars $Z,Y,X$.
The general problem for $p$ rows or columns was studied in \cite{Koch:2011hb} using a numerical approach. A key idea
was Schur-Weyl duality (also developed further in \cite{mn}) which enabled
a simple evaluation of the action of the dilatation operator.
For specific examples involving 3, 4 and 5 rows the diagonalization was
performed numerically, demonstrating a concrete connection to the   
Gauss Law constraints discussed in section 2.
Based on these numerical results, \cite{Koch:2011hb} conjectured the expression
(\ref{lovelyanswer}),  where integers $n_{ij}$ giving  the number of strings stretched 
between branes $i$ and $j$, appears in  a factored form of the action of 
the one-loop dilatation operator. In this paper, we prove this expression (\ref{lovelyanswer}).

In parallel developments, the problem of diagonalizing the free field 
inner product for  multi-matrix operators, in a way that preserves global symmetries 
was done in \cite{BHR1,BHR2}.  The group-theoretic construction
of these diagonal bases relied on the notion of Fourier transform 
on a finite group. It also showed the intimate relation between 
the counting of operators, refined according to global symmetries, 
and the actual construction of these operators. Often the counting, 
when expressed in the right group-theoretic language, provides 
natural hints for the actual construction of these operators. 
This theme was developed further in \cite{countconst} to
study eighth-BPS operators at weak coupling. At leading order in large $N$, 
these are just symmetrized traces made from three matrices $X , Y , Z $. 
A systematic procedure to construct  $1/N$-corrected BPS operators
was given using the permutation group algebra. This procedure 
found another use for the concept of ``counting to construction,''
whereby tools which give an elegant counting provide the necessary 
hints for the construction of the operators.  
Another relevant development appeared in   \cite{countFeyn}, where 
permutation group  methods for graph counting were reviewed 
and extended for various applications in counting Feynman graphs. 
Double cosets involving permutation groups played a significant role.

This paper starts with a general proof 
that the counting of states that can be constructed from 
restricted Schur operators matches the expectation from the Gauss Law.  
We focus on the case where there are $p$ (order one)  
giants, large either in the $AdS_5$  or the $S^5$, which 
are distinct. They have attached strings made of one type of 
building block, namely one impurity $Y$. 
We express the general counting of these brane-string configurations 
in terms graphs, which we call ``Gauss graphs.'' In this formulation, it 
becomes apparent that the number of these Gauss graphs is equal to 
the number of points in a double coset. The counting of these points 
is shown to be equal to the expected counting of operators in the restricted Schur construction. This is the first step of the 
{\it counting to construction philosophy} applied to these brane-string 
configurations. 

Fourier transformation applied to the double coset, 
gives a basis of functions, constructed from 
representation theory. This naturally leads to an explicit formula 
for the wavefunction in the $(s, \mu , \nu) $ sector. This wavefunction 
is labeled by elements of the double coset.
The full wavefunction is labeled by $ R$ and an element $\sigma $ 
of the double coset. The action of the one-loop dilatation operator 
takes the simple form (\ref{actiondil}).

Section 2 describes how the 
Gauss Law constraints, as applied to the string-brane configurations, 
lead to a graph counting problem. The result of this graph counting 
shows that the number of Gauss graphs is
equal to the number of points on a double coset
of permutation groups. The counting of the relevant restricted Schur 
polynomials is shown to match the size of this double coset, demonstrating 
that the physics of the Gauss Law for the compact branes correctly
matches the construction of operators by associating impurity insertions 
to the attached strings as conjectured by \cite{Balasubramanian:2004nb}. The  mathematical 
equivalence leading to the identity is related to Schur-Weyl duality, 
a theorem that has proved, in many instances,
to be a central instrument of  gauge-string 
duality \cite{ Cordes:1994fc,Ramgoolam:2008yr,Kimura:2008ac,Koch:2011hb,mn}.

Section 3 considers Fourier transformation on the double coset 
which appears in Section 2, and proposes Gauss graph operators
in $\cN=4$ SYM that utilize the Fourier coefficients that arise in the 
expansion of the delta function on the double coset. These Gauss graph 
operators are labeled by elements of the double coset. 
 
Section 4 proves that the  
one-loop dilatation operator acts diagonally in these 
double coset elements, to produce a differential 
operator acting on the $R$ label (\ref{lovelyanswer}). The structure of this 
differential operator as an element of $U(p)$ has
been previously recognized in \cite{gs} and is related to 
a system of $p$ particles in a line with 2-body harmonic oscillator
interactions.

Finally, a comment on notation is in order. In what follows, we will explicitly
indicate all sums over multiplicity labels and over representation labels.
For state labels we use the usual summation convention, that is, repeated\
indices are summed.

\section{Gauss Law : graphs and counting }

Our goal in this section is to argue that the number of states of an excited system of 
separated giant gravitons\footnote{None of the giant worldvolumes are coincident.}
is equal to the number of restricted Schur polynomials, labeled by Young diagrams with
widely separated corners.

\subsection{Restricted Schur Polynomials}

The restricted Schur polynomial is given by \cite{Balasubramanian:2004nb,Bhattacharyya:2008rb}
\bea\label{restschur}
  \chi_{R,(r,s)\mu\nu}(Z,Y)
  ={1\over n!m!}\sum_{\sigma\in S_{n+m}}{\rm Tr}_{R}\left( P_{R,(r,s)\mu\nu}\Gamma_R (\sigma)\right)
                        {\rm Tr}_{V_N^{\otimes\, n+m}}\left(\sigma\,Z^{\otimes\, n}Y^{\otimes\, m}\right)
\eea
where $R\vdash m+n$, $r\vdash n$ and $s\vdash m$. The operator 
$P_{R,(r,s)\mu\nu}$ is defined using 
an $S_n\times S_m$ irrep $(r,s)$ subduced by $R$, i.e. when the 
irrep $R$ of $S_{m+n}$ is decomposed into irreps of the $S_n \times S_m $
subgroup, $(r,s)$ is one of the irreps that appears in the decomposition. 
The labels $\mu , \nu $ run over the multiplicity with which 
$(r,s)$ appears in this restriction, a multiplicity which is 
equal to the Littlewood-Richardson number $g ( r ,s ; R )$.  
A basis of states in the irrep of $S_{m+n}$ corresponding to 
Young diagram $R$ can be given in terms of {\it standard tableaux} 
which are labelings of the Young diagram with 
integers $1$ to $m+n$ \cite{FH,Koch:2011hb,mn}. 
These integers in the standard tableaux
are strictly decreasing down the columns and 
along the rows.

Among these Young tableaux, if we consider all those 
that have the integers $1$ to $m$ entered in fixed locations, 
and the integers $\{ m+1 \cdots m+n \} $ in arbitrary locations, 
we get complete irreps of $S_n$. A useful way to 
think about this approach to the reduction from  
$S_{m+n}$ to $S_n$ is to use partially labeled Young 
tableaux 
\cite{Koch:2011hb,mn} where the remaining $n$ boxes are left unlabelled.
The unlabelled boxes determine a Young diagram  $r$ of $S_n$.  
The different partial labelings of the remaining boxes 
with $\{ 1, \cdots , m \} $ 
form the basis of a vector space which span the states in  irreps $s$ of $S_{m}$. Given the 
way $(n,m)$ appear in (\ref{restschur}),
 we may think of the $Y$'s as ``impurities'' 
which are replacing $Z$'s and correspondingly we may think of 
the labelings $\{ 1 , \cdots , m \} $ as specifying an order of 
removing ``$Y$-boxes'' from the ``$Z$-Young diagram'' $R$ to leave a
 $Z$-Young diagram $r$.

When the  $m$ $Y$ boxes are  thus assembled into an
irreps of $S_m$,  an  irrep  $s$  can occur with some multiplicity. 
The labels $\mu_1,\mu_2$ run over this multiplicity. Concretely we can write
\bea
  P_{R,(r,s)\mu_1\mu_2}={\bf 1}_r\otimes |s\,\mu_1\, ;\, i\rangle\,\langle s\, \mu_2\, ;\, i|
\label{proj}
\eea
where the $s$ state label $i$ is summed.

For restricted Schur polynomials corresponding to a system of $p$ giant gravitons we need $R$ to have
$p$ rows which each have $O(N)$ boxes. Further, for a system of separated giant gravitons none of the 
$p$ rows have the same length. We will focus on operators for which the row lengths in $R$ differ by
$O(N)$ boxes in the large $N$ limit. In this situation a concrete construction of the projectors (\ref{proj}) has been given in \cite{Koch:2011hb}. 
Each removed box is represented by a vector in
a $p$-dimensional vector space $V_p$.  Since we are removing $m$ boxes, 
the different ways of removing these span a vector space $V_p^{\otimes m }$. 
If the box $k$ is removed from row $i$, then the vector in the  $k$'th 
tensor factor has all zero entries except for the $i$th 
entry which is a 1. Introduce the vector $\vec{m}$ whose components
$m_i$ record the number of boxes removed from row $i$ of $R$ to produce $r$. 
The $m_i$ also correspond to the number of open strings emanating from the $i$th giant. Working with a basis of $V_p^{\otimes m } $, where the states 
have fixed $\vec m $, leads to the consideration of projectors
\bea
  P_{\vec m ; R,(r,s)\mu_1\mu_2}={\bf 1}_r\otimes |\vec{m}\, s\, \mu_1\, ;\,i\rangle\,\langle\vec{m}\, s\, \mu_2\, ;\, i|
\label{betterproj}
\eea
which represents a refinement of (\ref{proj}).  
In \cite{Koch:2011hb}
it was argued that when the corners of $R$ are well separated
the vector $\vec{m}$ is conserved by the dilatation operator 
so we can consider the action of dilatation operator on projectors 
of fixed $\vec m $. These conserved $U(1)^p$ charges will be explained 
more in section \ref{sec:dilop}. 

To count the number of restricted Schur polynomials it is useful to recall some facts about 
$V_p^{\otimes m}$ using Schur-Weyl duality. We will do this in section \ref{SW} below.

In what follows, it proves convenient to work with rescaled restricted Schur polynomials that have unit two 
point function. We denote the normalized operators by $O_{R,(r,s)\mu_1\mu_2}$.

\subsection{States Consistent with the Gauss Law}

A giant graviton has a compact world volume so that the Gauss Law implies the total charge on the giant's
world volume must vanish. Since the string end points are charged, this gives a constraint on the possible
open string configurations that are allowed: the number of strings emanating from the giant must equal the 
number of strings terminating on the giant.

\begin{figure}[h]
\begin{center}
\resizebox{!}{6cm}{\includegraphics{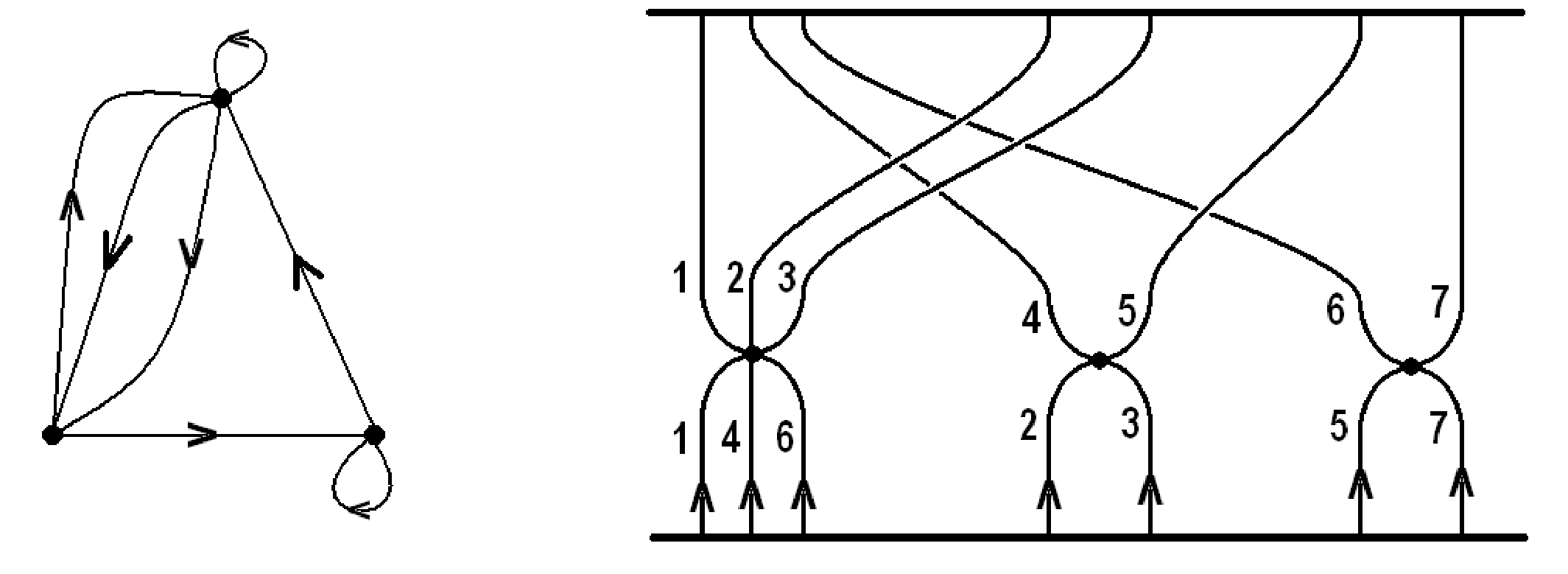}}
\caption{Any open string configuration can be mapped to a labeled graph as shown.
         The two bold horizontal lines are identified. 
         The graph itself determines a permutation, so each open string configuration
         is mapped to a permutation. 
         For the graph shown the permutation in cycle notation is $\sigma= (24)(536)$.
         As another example, the configuration in which all open strings loop back to the brane they
         start from is described by the identity permutation. 
         The figure shows a configuration for a three giant system with seven open strings attached.}
 \label{fig:gaussgraph}
\end{center}
\end{figure}

Each open string configuration corresponds to a graph, where the vertices 
represent the brane and the directed links represent oriented strings. 
Group theoretic graph counting techniques will be useful 
in counting these graphs (for a review and application
 to Feynman graphs in a variety of field theory problems see
 \cite{countFeyn} while some key earlier literature is \cite{Read}). 
To provide a systematic description of these
open string configurations, we describe the graphs using some numbers. 
Consider a case where there are a total of $m$ strings and $p$ branes.  
A convenient way to obtain a combinatoric description of the graphs we
consider is to divide each string into two halves and label 
each half.
Since the strings are oriented we can label the
outgoing ends with numbers $\{ 1, \cdots , m \} $ and the 
ingoing ends with these same numbers. 
 How the halves are joined is specified by a permutation $\sigma \in S_m$. 
Let $(m_1, m_2, \cdots , m_p)$ 
be the number of strings emanating from the distinct branes labeled from 
$1$ to $p$, so that $m_1 +m_2 + \cdots m_p = m$. 
 By the Gauss law, the numbers of strings ending at these branes 
is also given by the same ordered sequence of integers 
$(m_1, m_2 , \cdots , m_p)$. We can choose the labels of the 
half-strings such that the ones emanating from the first brane 
 are labeled $\{ 1 , 2 , \cdots , m_1 \} $, those emanating from 
the next set are labeled $\{ m_1 +1 , \cdots  m_2 \} $ etc.  Likewise 
the half-strings incident on the first brane are labeled  
$\{ 1 , 2 , \cdots , m_1 \} $, those incident on the second brane are labeled 
$ \{ m_1 +1 , \cdots  m_2 \} $ etc. The structure of the graph 
is encoded in the  permutation $\sigma \in S_m$ which describes 
how the $m$ outgoing half-strings are tied to the $m$ ingoing half-strings. 
There is some redundancy in this coding, because the $m_i$ strings 
emanating from  the $i$'th brane are indistinguishable, and likewise 
the $m_i$ strings incident on the $i$'th brane are indistinguishable. 
For an example of this labeling, see
the graph shown in figure \ref{fig:gaussgraph}. 
From the figure \ref{fig:gaussgraph} it is immediately clear that permutations 
which differ only by swapping end points that connect to the same
vertex do not describe distinct configurations. 
Relabeling of the outgoing half-strings, by permutations in their 
symmetry group $\prod_i S_{m_i}$,  acts on the permutation $\sigma$ 
describing the graph by a left multiplication, while relabeling the 
ingoing half-strings, by permutations in their symmetry group 
$\prod_i S_{m_i} $,  multiplies $\sigma$ on the right. 
The open string configurations are thus 
in one-to-one correspondence with elements of the double coset
\bea
  H \setminus S_m /H
\label{gaussdc}
\eea
where the group $H$ is $S_{m_1}\times S_{m_2}\times \cdots\times S_{m_p}$. 
This subgroup of $S_m$  will appear extensively in what follows.
Each element of the double coset gives a distinct graph of the type shown in Figure \ref{fig:gaussgraph}. We
call these ``{\it Gauss graphs}''.

Using the Burnside Lemma, the number of open string configurations $N_C$
(equivalently number of Gauss graphs)  is
\bea
  N_C ={1\over |H|^2 }
       \sum_{\alpha_1\in H}\sum_{ \alpha_2\in H}\sum_{\sigma_1\in  S_m}
       \delta (\alpha_2\sigma_1^{-1}\alpha_1^{-1}\sigma_1 )
 \label{ingap}
\eea
The delta function $\delta ( \alpha )  $ on the group is defined as $1$
if $\alpha $ is the identity and $0$ otherwise.  
We can rewrite this as
\bea
  N_C=
  {1\over |H|^2}
  \sum_{s\vdash m}\sum_{\alpha_1\in H}\sum_{ \alpha_2\in H} \chi_s(\alpha_2 )\chi_s(\alpha_1 )
\eea
The expression $ s\vdash m $ indicates that s is being summed 
over partitions of $m$, which describe Young diagrams corresponding to 
irreps of $S_m$. 
The sums over $\alpha_1$ and $\alpha_2$ produce projection operators which project onto the trivial representation 
of $H$.
Let $\cM^{ s }_{1_{H}}$ is the multiplicity of the one-dimensional representation 
of $H$ when the irreducible representation $s$ of $S_{m}$ is decomposed into representations of the subgroup $H$. 
The above formula is equivalent to 
\bea\label{countgauss}
N_C  = \sum_{ s \vdash m  } ( \cM^{ s }_{ 1_{H}})^2
\eea
We can also count using cycle indices
\bea
  N_C =N (Z(H) * Z(H) ) = \sum_{q\vdash m} Z_q^2 {\rm Sym}(q)
\eea
We know the cycle index of a product is the product of cycle indices so that
\bea
  Z(H)=\prod_i Z(S_{m_i})
\eea

\subsection{Two ways to decompose $V_p^{\otimes m }$  and refine by  $U(1)^p$ charges} 
\label{SW}

We can write 
\bea 
  V_p = \oplus_{ i=1}^p V_i 
\eea
The vector space $V_i$ is a one-dimensional space, spanned by the
eigenstate of $E_{ii}$ with eigenvalue one. If $v_i \in V_i$ then
\bea 
&& E_{ii} v_j = \delta_{ij}  v_i \cr 
&& E_{ij} v_k = \delta_{jk} v_i 
\eea
In the restricted Schur polynomial construction of \cite{Koch:2011hb} for long rows, 
a state in $V_i$ corresponds to a  $Y$-box in the $i$'th row. 

We have 
\begin{align}\label{decomp1} 
V_p^{\otimes m } & = \bigoplus_{ \substack{ s \vdash m  : \\ 
  c_1 ( s ) \le p }}  V_s^{ U(p)} 
\otimes V_s^{ S_m } \cr
& =  \bigoplus_{ \substack{ s \vdash m  : \\ c_1 ( s ) \le p }} 
 \bigoplus_{ \vec m } 
\otimes_{i=1}^m V_{m_i}^{U_i (1) }  \otimes V_{s\rightarrow \vec m }^{ U(p) \rightarrow U(1)^p }  \otimes V_s^{ S_m }
\end{align} 
Here $ \vec m $ is giving the $U(1)$ charges, with 
$\displaystyle{ \sum_{i=1}^p m_i = m }$. In the first 
line, we used Schur-Weyl duality. In the second, we 
decompose the $U(p)$ irrep $R$ into $U(1)^p $ irreps,
summing over all the irreps of this subgroup, labeled by $\vec m $. 
$V_{m_i}^{U_i (1) }$ is the one-dimensional irrep which transforms 
with charge $i$ under the $i$'th $U(1)$. In the restricted Schur
construction for long rows, these are the numbers of boxes in the $i$'th row. 
Each set of $U(1)$ charges $\vec m $ will come with a multiplicity 
label. These multiplicity labels span a vector space  $V_{s\rightarrow \vec m }^{ U(p) \rightarrow U(1)^p }$. The dimension of that vector space 
is the number of times the irrep $\vec m $ of $U(1)^p$ appears when the 
irrep $s$ of $U(p)$ is decomposed under the subgroup  $U(1)^p$. 
These are the Kotska numbers \cite{FH} denoted by  $ {\cal K}_{ s\, \vec{m}}$. 
Since the restricted Schur polynomials are labeled by a pair of multiplicity
labels, the total number of restricted Schurs is the sum of the squares
of the Kostka numbers
\bea\label{countrest}
\hbox { Number of restricted Schur polynomials } = \sum_{\substack{ s \vdash m  : \\  c_1 ( s ) \le p }} ( {\cal K}_{ s\, \vec{m}})^2
\eea
The goal of this section is to prove the equality of the number of configurations consistent with the
Gauss Law, given by (\ref{countgauss}), and the number of restricted Schurs, given by (\ref{countrest}).
This equality is a consequence of Schur-Weyl duality, which we now develop more fully.

We can  develop the steps above at the level of a 
basis for $ V_p^{\otimes m} $. The reduction coefficients 
that will arise in the final step are the branching coefficients 
for irrep $s$ of $U(p)$ into the irrep $\vec m $ of $ \cH \equiv  U(1)^p  $. 
Indeed we can write $ | I \rangle $ as a shorthand for the tensor basis
$ | i_1 , i_2 , \cdots , i_p \rangle $. From Schur-Weyl duality, we know 
there is a change of basis to 
\bea 
| I \rangle = \sum_{ s , m_s , M_s } 
| s , M_s , m_s \rangle  \langle s , M_s , m_s | I \rangle  
\eea
The label $M_s$ is a state label for the $U(p)$ irrep $s$. 
It corresponds to semi-standard Young tableaux, as reviewed 
in Appendix A of \cite{Koch:2011hb}. The label $m_s$, a state label 
for the $S_m$ irrep $s$, can be described by standard Young tableaux. 
We can now decompose into $U(1)^p$ 
\bea 
| I \rangle = \sum_{ \vec m , \nu }  \sum_{ s , m_s , M_s } 
C_{ M_s   }^{ \vec m , \nu  }   | s , \vec m  , m_s \rangle  \langle s , M_s , m_s | I \rangle  
\eea 
The coefficient $C_{ M_s   }^{ \vec m , \nu  } $ 
gives the decomposition of a $U(p)$ irrep into $U(1)^p$ 
irreps, and contains a multiplicity label $\nu$. This
multiplicity label is labeling states in 
$V_{s\rightarrow \vec m }^{ U(p) \rightarrow U(1)^p }$.

There is an alternative way to decompose $V_p^{\otimes m }$ 
into irreps of $\cH = U(1)^p $ by using permutations in $S_m$. Observe that when we 
choose charges $\vec m $, then there are $m_1$ copies of 
$v_1 $ , $m_2 $ copies of $v_2$ etc. One such state is 
\bea\label{defvbar}  
| \bar v , \vec{m}\rangle \equiv | 
v_1^{\otimes m_1 } \otimes v_2^{\otimes m_2 } \otimes \cdots  v_p^{\otimes m_p }
\rangle
\eea
A general state with these charges 
can be obtained by a permutation of the above. 
\bea 
| v_{\sigma } \rangle  \equiv \sigma | v_1^{\otimes m_1 } \otimes v_2^{\otimes m_2 } \otimes \cdots  v_p^{\otimes m_p }
\rangle
\eea
where 
\bea 
\sigma | v_{i_1} \otimes \cdots \otimes v_{i_p } \rangle 
=  | v_{i_{ \sigma (1)} }\otimes  \cdots \otimes v_{i_{ \sigma( p)  } } \rangle 
\eea
Clearly not all $\sigma $ give independent vectors 
\bea 
| v_{ \sigma } \rangle = | v_{ \sigma \gamma } \rangle
\eea
if $ \gamma \in H $.

We can write 
\bea 
| v_{ \sigma } \rangle  = { 1 \over | H| } \sum_{ \gamma \in H} 
|  v_{ \sigma \gamma } \rangle
\eea
In other words the states are in correspondence with $S_{m}/ H$.
A convenient description of these states can be developed
using representation theory,
exploiting methods of \cite{BHR1,BHR2}. 
Look at the representation basis
\bea 
| v_{ s , i ,  j }\rangle 
 && = \sum_{ \sigma\in S_m } \Gamma^{(s)}_{ij} ( \sigma ) | v_{ \sigma} \rangle \cr 
&& ={ 1 \over |H| }   \sum_{ \sigma\in S_m } \sum_{ \gamma \in H }
 \Gamma^{(s)}_{ij} ( \sigma  ) | v_{ \sigma \gamma } \rangle \cr
&& =  { 1 \over |H| }   \sum_{ \sigma\in S_m } \sum_{ \gamma \in H }
 \Gamma^{(s)}_{ij} ( \sigma \gamma  ) | v_{ \sigma  } \rangle \cr
&& =   { 1 \over |H| }  \sum_{ \sigma\in S_m } \sum_{ \gamma \in H }
 \Gamma^{(s)}_{ik} ( \sigma )  
\Gamma^{(s)}_{kj } ( \gamma  ) | v_{ \sigma  } \rangle \cr 
&& = \sum_{ \sigma\in S_m } \sum_{ \mu  }  \Gamma^{(s)}_{ik} ( \sigma ) 
B^{s \rightarrow 1_H}_{ k \mu}  B^{s \rightarrow 1_H}_{ j \mu}   | v_{ \sigma  } \rangle
\eea
In the last line above
we have decomposed the matrix elements of the $H$ projector 
into products of branching coefficients using
\bea
{1\over |H|}\sum_{ \gamma \in H } \Gamma^{(s)}_{ik} ( \sigma )
=\sum_\mu B^{s \rightarrow 1_H}_{ i \mu}  B^{s \rightarrow 1_H}_{ k \mu}
\eea
It is now natural to introduce
\bea\label{u1pdec}  
 | \vec m  , s, \mu ; i \rangle \equiv  \sum_j B^{s \rightarrow 1_H}_{ j \mu}
 | v_{ s , i ,  j }\rangle =  \sum_j B^{s \rightarrow 1_H}_{ j \mu}  \sum_{ \sigma\in S_m } \Gamma^{(s)}_{ij} ( \sigma ) | v_{ \sigma} \rangle
\eea
In this construction, the $\mu$ index is a multiplicity 
for reduction of $S_m$ into $H$
and the group-theoretic transformations $B^{s \rightarrow 1_H}_{ j \mu} $ involved have to do with 
$S_m \rightarrow H$. In the construction earlier we had 
$C_{ M_s   }^{ \vec m , \nu  }  $ associated to $U(p) \rightarrow U(1)^p$, 
which are closer to Gelfand-Tsetlin bases used in \cite{Koch:2011hb}.

We can now prove  the equality of Kotska numbers
(defined in terms of reduction multiplicities of $U(p)$ to $U(1)^p$)
and the branching multiplicity of $S_m \rightarrow H$. 
The decomposition of $V_p^{ \otimes m } $ refined according to $U(1)^p$
in the second way we have done it is 
\bea\label{decomp2}  
V_p^{\otimes m } = \bigoplus_{ \vec m } \bigoplus_{ s} V_s^{S_m } \otimes 
V_{ s \rightarrow {\bf 1}  }^{ S_m \rightarrow H(\vec m )  }  \otimes_{i=1}^p V_{m_i}^{U(1)^p }   
\eea
Compare (\ref{decomp1}) to (\ref{decomp2})  to deduce 
\bea 
\cM^{ s }_{ {\bf 1}_{H}} \equiv
|V_{ s \rightarrow {\bf 1}  }^{ S_m \rightarrow H(\vec m )  } | =   |V_{s\rightarrow \vec m }^{ U(p) \rightarrow U(1)^p } | 
\equiv {\cal K}_{ s\, \vec{m}}
\eea
which is the desired equality between Kotska numbers for $U(p) \rightarrow U(1)^p$
and branching multiplicities for $S_{m} \rightarrow H $. 
This completes the proof of the equality between (\ref{countgauss}) and (\ref{countrest}).

\section{Gauss Operators}

In the previous section we argued that Gauss graphs are described by elements of the double coset (\ref{gaussdc}). In a number of problems related to the 
construction of gauge-invariant operators in the context of gauge-string 
duality, it is found that counting results for gauge invariant operators, 
once expressed in appropriate group theoretic language, 
lead naturally to methods for the explicit construction of these operators.
This occurs notably in the study of eighth-BPS operators at zero Yang-Mills
 coupling, which involves diagonalizing the free field inner product 
for holomorphic gauge-invariant multi-matrix operators \cite{dolan,rajesh,BHR1,BHR2}. The link between counting to construction 
often involves Fourier transforms on groups.  
This counting to construction philosophy was developed further in 
 \cite{countconst} in the context of eighth  BPS operators at weak coupling. 
We may expect therefore that the double coset we have used to count the Gauss 
graphs should  also play an
important role in constructing the operators dual to the Gauss graph configurations. In this section we will
construct a complete set of functions on the double coset,
which give, as in usual Fourier analysis, 
 an expansion for the delta function, in this case,  on the double coset. 
This gives a  natural guess for the operators dual to a given Gauss graph configuration. 
In the next section we will see that one loop dilatation operator 
acts diagonally on the operators labeled by these double coset elements
which thus provide the diagonalization of the one-loop dilatation operator 
action on  $s,\mu , \nu $ labels of the restricted Schur operators $\cO_{R, r,s, \mu , \nu }$. 
This gives an analytic confirmation of the numerical results 
obtained in  \cite{Koch:2011hb} 
as well as a significant extension of these results to the general case.

The methods of representation theory used in this section have been 
used in the context of AdS/CFT for diagonalizing 
the free field inner product for multi-matrix operators
\cite{BHR1,BHR2}. Recall that
the matrix elements of irreducible representations $s \vdash m$ give a basis of functions on $S_m$. Given an object $\cO_{\tau } $ 
determined by a permutation $\tau$, we can form linear combinations 
$\cO^s_{ij}$ labeled by an irrep $s$ and state labels $i,j$. 
\bea 
\cO^s_{ij} =  \sum_{ \sigma \in S_m } \Gamma^{(s)}_{ij} ( \sigma ) \cO_{\sigma }  
\eea
This is an isomorphic description, which is not surprising
given the familiar group theory identity $m! = \sum_s d_s^2 $. 
Indeed these matrix elements provide a resolution of the delta-function 
on the group since 
\bea 
\sum_s { d_s \over m! } \Gamma^{(s)}_{ij} ( \sigma )  \Gamma^{(s)}_{ij } ( \tau  ) 
= \delta ( \sigma \tau^{-1} ) 
\eea
and indeed behave like Fourier coefficients.

Suppose we have some object determined by a permutation 
$\tau\in S_m$, call it $\cO_{\tau } $, but which is invariant 
under left and right multiplication of $\tau $ by $\gamma_1 , \gamma_2$
in the subgroup $H$. Here $H= H(\vec m ) = \prod_i S_{m_i}$. 

We can write 
\bea
 \cO_{\tau } && = { 1 \over|H|^2 } \sum_{ \gamma_1 , \gamma_2 \in H } 
               \cO_{\gamma_1 \tau \gamma_2 } \cr 
&& = { 1 \over |H|^2} \sum_s { d_s \over m! } \sum_{ \gamma_1 , \gamma_2 } 
     \Gamma^{(s)}_{ij} ( \gamma_1 \tau \gamma_2 ) \cO^s_{ij} \cr 
&& =   { 1 \over |H|^2} \sum_s { d_s \over m! } \sum_{ \gamma_1 , \gamma_2 }
     \Gamma^{(s)}_{ik} ( \gamma_1 )   \Gamma^{(s)}_{k l } ( \tau )  \Gamma^{(s)}_{ l j  } ( \gamma_2  )  
     \cO^s_{ij} \cr 
&& = \sum_s { d_s \over m! } \Gamma^{(s)}_{kl} ( \tau ) B^{s \rightarrow 1_H} _{i \mu_1 } B^{s \rightarrow 1_H} _{k \mu_1} 
                            B^{s \rightarrow 1_H} _{l \mu_2 } B^{s \rightarrow 1_H} _{j \mu_2 } \cO^s_{ij} \cr 
&& = \sum_s  
\left ( \sqrt { d_s \over m! }      
 \Gamma^{(s)}_{kl} ( \tau ) B^{s \rightarrow 1_H} _{k \mu_1 } B^{s \rightarrow 1_H} _{k \mu_2 } \right ) 
 \left (  \sqrt { d_s \over m! } B^{s \rightarrow 1_H} _{i \mu_1 } B^{s \rightarrow 1_H} _{j \mu_2 } \cO_{ij}^s \right )
\cr 
&& = \sum_{ s } 
\left ( \sqrt { d_s  \over m! } \Gamma^{(s)}_{kl} ( \tau ) B^{s \rightarrow 1_H} _{ k \mu_1} B^{s \rightarrow 1_H} _{l \mu_2 } \right ) \cO^s_{\mu_1 \mu_2 } 
\eea 

We have introduced branching coefficients for the 
trivial irrep of $H$ inside the representation $s$ of $S_m$. 
These $B^{s \rightarrow 1_H} _{i \mu }$ give the expansion 
of the $\mu$'th occurrence of the identity irrep of $H$ when
irrep $s$ of $S_m$ is decomposed into irreps of the subgroup $H$, 
in terms of the states labeled $i $ in $s$. We also defined 
the linear combinations 
\bea 
 \cO^s_{\mu_1 \mu_2 }  = \sqrt { d_s \over m! } B^{s \rightarrow 1_H} _{i \mu_1 } B^{s \rightarrow 1_H} _{j \mu_2 } \cO_{ij}^s
\eea
 labeled by the irrep label $s$ and a multiplicity label for the 
decomposition to the identity irrep of $H$. 
These provide the representation theoretic basis for the 
double coset in accordance with (\ref{countgauss}).

We now show that the group-theoretic coefficients 
\bea 
C_{\mu_1 \mu_2 }^{s }  ( \tau )  = 
|H|  \sqrt { d_s  \over m! } 
\Gamma^{(s)}_{kl} ( \tau ) B^{s \rightarrow 1_H} _{ k \mu_1} B^{s \rightarrow 1_H} _{l \mu_2 } 
\eea 
provide an orthogonal   transformation between double coset elements $\sigma $ 
and the  $ \cO^s_{  \mu_1 , \mu_2 }$.
The introduction of the normalization $|H|$ is for convenience. 
We can show that 
\bea 
C_{ \mu_1 \mu_2}^s ( \tau ) C_{\mu_1 \mu_2 }^s ( \sigma ) 
&& = |H|^2  \sum_{ s } { d_s \over m! }  B^{s \rightarrow 1_H} _{ k \mu_1 } B^{s \rightarrow 1_H} _{l \mu_2 }
     \Gamma^{(s)}_{kl} ( \tau ) 
     B^{s \rightarrow 1_H} _{ p \mu_1 } B^{s \rightarrow 1_H} _{ q \mu_2 } \Gamma_{pq}^{(s)} ( \sigma ) \cr 
&& = \sum_s \sum_{ \gamma_1 , \gamma_2 } 
   { d_s \over m! } \Gamma^{(s)}_{k p } ( \gamma_1 )   \Gamma^{(s)}_{l q  } ( \gamma_2 )
   \Gamma^{(s)}_{kl} ( \tau )  \Gamma^{(s)}_{p q } ( \sigma  ) \cr 
&& = \sum_s {d_s\over m!} \chi_s ( \gamma_1 \sigma \gamma_2^{-1} \tau^{-1} ) \cr 
&& = \sum_{ \gamma_1 , \gamma_2 } \delta ( \gamma_1 \sigma \gamma_2 
\tau^{-1} )   
 \eea
 
This expresses orthogonality since the right hand side 
is a delta function on the double coset, and shows that 
a representation theoretic way of counting 
the number of elements in the double coset is
\bea 
\sum_{s  } (\cM^{ s }_{ 1_H} )^2
\eea
in agreement with (\ref{countgauss}), which we previously obtained by 
applying the Burnside Lemma.

In view of this discussion, a very natural form for the operators dual to Gauss configuration $\sigma$, up to normalization, is
\bea
  O_{R,r}(\sigma)
  ={|H|\over \sqrt{m!}}\sum_{j,k}\sum_{s\vdash m}\sum_{\mu_1,\mu_2}\sqrt{d_s}
  \Gamma^{(s)}_{jk}(\sigma )B^{s\to 1_H}_{j \mu_1}B^{s\to 1_H}_{k \mu_2} O_{R,(r,s)\mu_1\mu_2}
  \label{ggo}
\eea
The overall factor has been chosen to ensure a convenient normalization.
Indeed, the two point function of Gauss graph operators is
\bea
\langle O_{R,r}(\sigma_1)O_{T,t}^\dagger (\sigma_2)\rangle
&&={|H|^2\over m!}\sum_{s,u\vdash \, m}\sum_{\mu_1\mu_2\nu_1\nu_2}\sqrt{d_s d_u}
\Gamma^{(s)}_{jk}(\sigma_1) B_{j\mu_1}^{s\to 1_H}B_{k\mu_2}^{s\to 1_H} \times\cr
&&\times \Gamma^{(u)}_{lm}(\sigma_2) B_{l\nu_1}^{u\to 1_H}B_{m\nu_2}^{u\to 1_H}\,
\langle O_{R,(r,s)\mu_1\mu_2}O_{T,(t,u)\nu_1\nu_2}^\dagger\rangle
\eea
Now, use (see Appendix \ref{nutshell})
\bea
\langle O_{R,(r,s)\mu_1\mu_2}O_{T,(t,u)\nu_1\nu_2}^\dagger\rangle
    =\delta_{rt} \delta_{su}\delta_{\mu_1\nu_1}\delta_{\mu_2\nu_2}
\eea
to obtain
\bea
\langle O_{R,r}(\sigma_1)O_{T,t}^\dagger (\sigma_2)\rangle 
&&={|H|^2 \over m!}\sum_{s\vdash \, m} d_s
\Gamma^{(s)}_{jk}(\sigma_1) B_{j\mu_1}^{s\to 1_H}B_{k\mu_2}^{s\to 1_H}
\Gamma^{(s)}_{lm}(\sigma_2) B_{l\mu_1}^{s\to 1_H}B_{m\mu_2}^{s\to 1_H}\cr
&&={1\over m!}\sum_{s\vdash \, m} \sum_{\gamma_1,\gamma_2\in H} d_s
\Gamma^{(s)}_{jk}(\sigma_1)\Gamma^{(s)}_{jl}(\gamma_1)
\Gamma^{(s)}_{lm}(\sigma_2) \Gamma^{(s)}_{mk}(\gamma_2)\cr
&&={1\over m!}\sum_{\gamma_1,\gamma_2\in H}\sum_s d_s
\chi_s (\sigma_1^{-1}\gamma_1\sigma_2\gamma_2)\cr
&&= \sum_{\gamma_1,\gamma_2\in H}\delta (\sigma_1^{-1}\gamma_1\sigma_2\gamma_2)
\label{cosettwopoint}
\eea
The right hand side is the delta function on the double coset, setting $\sigma_1 = \sigma_2$. 
Thus if $\sigma_1$ and $\sigma_2$ represent the same double coset element, the two point function is one and
if they represent distinct coset elements, it vanishes.

\section{Dilatation Operator}\label{sec:dilop} 

In this section we will review the exact action of the one loop dilatation operator on restricted Schur 
polynomials \cite{DeComarmond:2010ie}.
We then review how this action simplifies when acting on restricted Schurs with long rows and well 
separated corners \cite{Carlson:2011hy,Koch:2011hb}.
Using this simplified action we prove that the Gauss 
graph operators diagonalize 
the dilatation operator's $Y$ labels.

\subsection{Action of the Dilatation Operator}

When acting on restricted Schurs the one loop dilatation operator takes 
the form \cite{DeComarmond:2010ie}
$$
  DO_{R,(r,s)\mu_1 \mu_2}(Z,Y)=\sum_{T,(t,u)\nu_1\nu_2} N_{R,(r,s)\mu_1 \mu_2;T,(t,u)\nu_1\nu_2}O_{T,(t,u)\nu_1\nu_2}(Z,Y)
$$
where
\begin{eqnarray}
\label{dilat}
N_{R,(r,s)\mu_1 \mu_2;T,(t,u)\nu_1\nu_2}&&\!\!\!\!\!\!\!\!= - g_{YM}^2\sum_{R'}{c_{RR'} d_T n m\over d_{R'} d_t d_u (n+m)}
\sqrt{f_T \, {\rm hooks}_T\, {\rm hooks}_r \, {\rm hooks}_s \over f_R \, {\rm hooks}_R\, {\rm hooks}_t\, {\rm hooks}_u}\times
\\
\nonumber
&&\!\!\!\!\!\!\!\!\!\!\!\!\!\!\!\!\!\!\!\!\!\!\!\!\!\!\!\!\!\!\!\!\!\!\!\!\!\!\!\!\!\!
\times{\rm Tr}\Big(\Big[ \Gamma^{(R)}((1,m+1)),P_{\vec{m};R,(r,s)\mu_1 \mu_2}\Big]I_{R'\, T'}
\Big[\Gamma^{(T)}((1,m+1)),P_{\vec{m};T,(t,u)\nu_2\nu_1}\Big]I_{T'\, R'}\Big) \, .
\end{eqnarray}
The trace above is over the direct sum 
representation $R \oplus T $ where $ R , T $ are 
Young diagrams with $m+n$ boxes. 
$R'$ is one of the irreps subduced from $R$ upon restricting 
to the $S_{n+m-1}$ subgroup of $S_{n+m}$ obtained by keeping
only permutations that obey $\sigma (1)=1$. $T'$ is subduced by $T$ in the same way. $I_{R'\, T'}$ is an intertwining map
(see Appendix  D of \cite{Koch:2011hb} for details on its properties) 
 from irrep $R'$ to irrep $T'$.  It  is only non-zero if $R'$ and $T'$ have the same shape.
Thus, to get a non-zero result we need $R=T$ or $R$ and $T$ must differ at most by the placement of a single box.
$d_a$ denotes the dimension of symmetric group irrep $a$.
$f_S$ is the product of the factors in Young diagram $S$ and 
${\rm hooks}_S$ is the product of the hook lengths of Young diagram $S$.
Finally, $c_{RR'}$ is the factor of the corner box that must be removed from $R$ to obtain $R'$.

When acting on Schurs labeled by Young diagrams $R$ with long rows and well separated corners, it is possible to
compute $N_{R,(r,s)\mu_1 \mu_2;T,(t,u)\nu_1\nu_2}$ explicitly \cite{Carlson:2011hy,Koch:2011hb}. 
We will now review the relevant steps in this evaluation.
We consider $n\gg m$ and assume that $R$ has $p$ long rows.
We hold $p$ fixed and order 1 as we take $N\to\infty$.
In this limit the difference in the lengths of the corresponding rows of $R$ and $r$ can be neglected.

In the construction of the projectors we removed $m$ boxes from $R$ to produce $r$ with each box represented by a vector in $V_p$.
To evaluate the action of the dilatation operator, it is convenient to remove $m+1$ boxes again associating each with a vector in $V_p$.
This allows a straight forward evaluation of the action of $\Gamma^{(R)}\big( (1,m+1)\big)$ and $\Gamma^{(T)}\big( (1,m+1)\big)$. 

As mentioned above, $R$ and $T$ agree after removing a single box. 
The $R'$ and $T'$ subspaces are obtained by removing this single box from $R$ and $T$ respectively.
To produce a map from $R'$ to $T'$ we simply need a map from the vector corresponding to the box removed from $R$ to
the vector corresponding to the box removed from $T$. 
This map is $E^{(1)}_{ij}$ if we remove the box from row $i$ of $R$ and row $j$ of $T$.
Using the identification
\bea
  (1,m+1) = {\rm Tr}(E^{(1)}E^{(m+1)})
\eea
We easily find, for example (repeated indices are summed)
\bea
  E^{(1)}_{ji }\Gamma^{(R)}\left( (1,m+1)\right) = E^{(1)}_{ji}E^{(1)}_{kl}E^{(m+1)}_{lk}=E^{(1)}_{jl}E^{(m+1)}_{li}
\eea
An easy way to understand this result is to recognize that $E^{(1)}_{ji}=E^{(1)}_{ji}E^{(m+1)}_{ll}$ so that
$\Gamma^{(R)}\left( (1,m+1)\right)$ simply swapped the column labels. 
This simple action is a direct consequence of the simplified action of the symmetric group when the corners of $R$ are well separated.
After performing these manipulations we are left with a trace over products of $E_{ij}$s acting in slots 1 and $m+1$ and the
operators $P_{\vec{m};R,(r,s)\mu_1\mu_2}$ and $P_{\vec{n};T,(t,u)\nu_2\nu_1}$. The trace thus factorizes into a trace over irrep $r$ and a
trace over $V_p^{\otimes m}$. After performing these traces we have
\bea
  D O_{R,(r,s)\mu_1\mu_2}=-g_{YM}^2\sum_{u\nu_1\nu_2}\sum_{i<j}\delta_{\vec{m},\vec{n}}M^{(ij)}_{s\mu_1\mu_2 ; u\nu_1\nu_2}\Delta_{ij}
                           O_{R,(r,u)\nu_1\nu_2}
  \label{factoredD}
\eea
where $\Delta_{ij}$ acts only on the Young diagrams $R,r$ and
\bea
  M^{(ij)}_{s\mu_1\mu_2 ; u\nu_1\nu_2}&&={m\over\sqrt{d_s d_u}}\left(
       \langle \vec{m},s,\mu_2\, ;\, a|E^{(1)}_{ii}|\vec{m},u,\nu_2\, ;\, b\rangle
       \langle \vec{m},u,\nu_1\, ;\, b|E^{(1)}_{jj}|\vec{m},s,\mu_1\, ;\, a\rangle\right.\cr
&&\left. +
      \langle \vec{m},s,\mu_2\, ;\, a|E^{(1)}_{jj}|\vec{m},u,\nu_2\, ;\, b\rangle
      \langle \vec{m},u,\nu_1\, ;\, b|E^{(1)}_{ii}|\vec{m},s,\mu_1\, ;\, a\rangle\right)
\eea
where $a$ and $b$ are summed. $a$ labels states in irrep $s$ and $b$ labels states in irrep $t$. The action of operator
$\Delta_{ij}$ is most easily split up into three terms
\bea
  \Delta_{ij}=\Delta_{ij}^{+}+\Delta_{ij}^{0}+\Delta_{ij}^{-}
\eea
Denote the row lengths of $r$ by $r_i$. 
The Young diagram $r_{ij}^+$ is obtained by removing a box from row $j$ and adding it to row $i$.
The Young diagram $r_{ij}^-$ is obtained by removing a box from row $i$ and adding it to row $j$.
In terms of these Young diagrams we have
\bea
  \Delta_{ij}^{0}O_{R,(r,s)\mu_1\mu_2} = -(2N+r_i+r_j)O_{R,(r,s)\mu_1\mu_2}
  \label{0term}
\eea
\bea
  \Delta_{ij}^{+}O_{R,(r,s)\mu_1\mu_2} = \sqrt{(N+r_i)(N+r_j)}O_{R^+_{ij},(r^+_{ij},s)\mu_1\mu_2}
  \label{pterm}
\eea
\bea
  \Delta_{ij}^{-}O_{R,(r,s)\mu_1\mu_2} = \sqrt{(N+r_i)(N+r_j)}O_{R^-_{ij},(r^-_{ij},s)\mu_1\mu_2}
  \label{mterm}
\eea
Notice that $\Delta_{ij}$ acts on $r$ i.e. on $Z$s and $M^{(ij)}_{s\mu_1\mu_2 ; u\nu_1\nu_2}$ on $Y$s.
Note that it is a consequence of the fact that $R$ and $r$ change in exactly the same way that $\vec{m}$ is preserved
by the dilatation operator. As a matrix $\Delta_{ij}$ has matrix elements
\bea
  \Delta^{R,r ;  T,t}_{ij} =
 \sqrt{(N+r_i)(N+r_j)}(\delta_{T,R^+_{ij}}\delta_{t,r^+_{ij}}+
\delta_{T,R^+_{ij}}\delta_{t,r^+_{ij}})
                                          -(2N+r_i+r_j)\delta_{T,R}\delta_{t,r}
\eea
In terms of these matrix elements we can write (\ref{factoredD}) as
\bea
  D O_{R,(r,s)\mu_1\mu_2}=-g_{YM}^2\sum_{T,(t,u)\nu_1\nu_2}\sum_{i<j}\delta_{\vec{m},\vec{n}}M^{(ij)}_{s\mu_1\mu_2 ; u\nu_1\nu_2}\
                            \Delta^{R,r ; T,t}_{ij}   O_{T,(t,u)\nu_1\nu_2}
  \label{factoredDsecond}
\eea

\subsection{Diagonalization}

Given the factorized dilatation operator (\ref{factoredD}), we can diagonalize on the $s\mu_1\mu_2 ; u\nu_1\nu_2$
and the $R,r ; T,t$ labels separately. In this section we are mainly concerned with describing the result of diagonalizing
on the $s\mu_1\mu_2 ; u\nu_1\nu_2$ labels. This result was obtained analytically for two rows. For more than two rows the
results are numerical, motivating a conjecture we describe in this section. In the next section we will provide an analytic 
treatment valid for any number of rows, thereby proving the conjecture.

After diagonalization on the $s\mu_1\mu_2 ; u\nu_1\nu_2$ labels one obtains a collection of decoupled eigenproblems
in the $R,r ; T,t$ labels. There is one eigenproblem for each Gauss graph that can be drawn and the structure of each 
problem is naturally read from the Gauss graph. To obtain the problem associated to a particular Gauss graph, count 
the number $n_{ij}$ of strings (of either orientation) stretching between branes $i$ and $j$. For example, the Gauss
graph of Figure \ref{fig:gaussgraph} has $n_{12}=1$, $n_{13}=3$ and $n_{23}=1$. The action of the dilatation operator
on the Gauss graph operator is
\bea\label{actiondil} 
   DO_{R,r}(\sigma) = -g_{YM}^2 \sum_{i<j}n_{ij} ( \sigma ) 
\Delta_{ij} O_{R,r}(\sigma)
\eea
To obtain anomalous dimensions one needs to solve an eigenproblem on the $R,r$ labels.
We have anticipated the fact that it is the Gauss graph operators defined above that accomplish this diagonalization.
This is one of the key results of this article and will be proved in the next section. Towards this end, it is useful 
to develop a formula for $n_{ij}$ in terms of $\sigma$.

For $i  < j $, let $n^+_{ij}$ be the number of strings going from $i$ to $j$ and $n^-_{ij} $ the number from $j$ to $i$. 
Since $n_{ij}$ is orientation blind we have $n_{ij}=n^+_{ij} + n^-_{ij}$.
If $k$ is in the range $ \{  m_1 + \cdots + m_{i-1} +1 , \cdots , m_1 + \cdots + m_i \}  $, then $n_{ij}^{+} $ is the number of $ \sigma (i ) $ lying in the range $ \{   m_1 + \cdots + m_{j-1} + 1  , \cdots ,  m_1 + \cdots +  m_{j } \} $. 
\bea 
   n_{ij}^+  ( \sigma ) = \sum_{ k = m_1 + \cdots + m_{i-1} +1  }^{
   m_1 + \cdots + m_{i}  }  ~~~~  \sum_{ l  = m_1 + \cdots + m_{j-1} +1  }^{
   m_1 + \cdots + m_{j}  } \delta ( \sigma(k  ) , l )   
\eea

Equivalently if we say that $S_1, S_2 , \cdots , S_p$ are, respectively, 
the first $m_1$ positive integers, the next $m_2$, and so on, 
then 
\bea 
   n_{ij}^{ + } ( \sigma ) = \sum_{ k \in S_i } \sum_{ l \in S_j }
   \delta ( \sigma (k) , l ) 
\eea
 
Similarly the number of strings going the other way is 
\bea 
   n_{ij}^{ - } ( \sigma ) = \sum_{ k  \in S_i } \sum_{ l \in S_j }
   \delta ( \sigma (l) , k ) 
\eea

\begin{figure}[h]
\begin{center}
\resizebox{!}{6cm}{\includegraphics{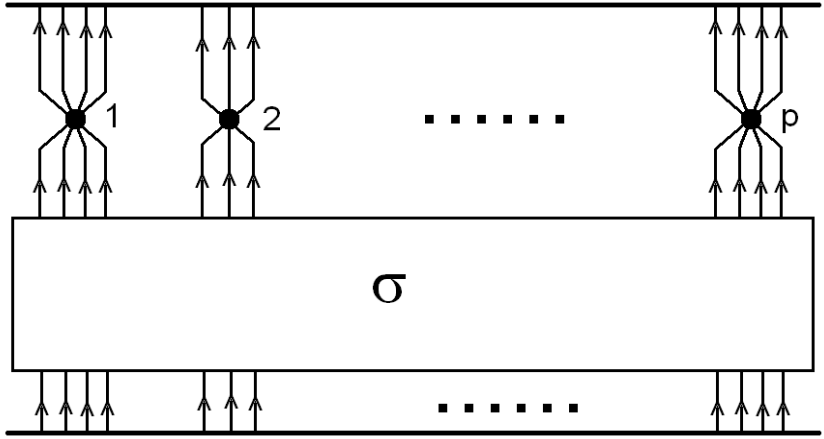}}
\caption{The number of open strings emanating on each brane is described by $\vec{m}$. The permutation $\sigma$
         specifies how these strings are to be terminated on the branes.}
 \label{fig:doublecoset}
\end{center}
\end{figure}

\subsection{Action on Gauss graph operators}

Having defined the  kets $|\vec{m}, s , \mu ; i\rangle $, 
we will now think about the bras $ \langle \vec{m}, u, \nu ; j |$. The following definition 
\bea 
\langle  \vec{m} , u, \nu ; j|  =  {  d_u\over m! |H|  }  \sum_{ \tau \in S_m } 
\langle \bar v ,\vec{m} | \tau^{-1} \Gamma^{(u)}_{j k } ( \tau ) B^{u\to 1_H}_{k\nu }  
\eea
will the give correctly normalized relation
\bea 
\langle \vec{m}, u, \nu ; j |\vec{n} , s , \mu ; i\rangle = \delta_{\vec{m}\vec{n}} \delta_{us} \delta_{ji} 
\delta_{\mu\nu} 
\eea
To see this calculate 
\bea 
\langle  \vec{m} , u, \nu ; j| \vec{n} , s , \mu ; i \rangle 
&& = {  d_u \over m! |H|  }  \sum_{ \tau } 
\sum_{ \sigma } \langle \bar v ,\vec{m} | \tau^{-1} 
\Gamma^{(u)}_{  j k  } ( \tau ) B^{u\to 1_H}_{k\nu}\Gamma^{(s)}_{ i l }    ( \sigma  ) 
 B^{s\to 1_H}_{l\mu } | \bar v ,\vec{n}\rangle \cr 
&&= {  d_u \over m! |H|} \sum_{ \tau , \sigma } \Gamma^{(u)}_{  j k  } ( \tau ) 
 B^{u\to 1_H}_{k\nu}  \Gamma^{(s)}_{ i l }    ( \sigma  ) 
 B^{s\to 1_H}_{l\mu } \langle \bar v ,\vec{m}| \tau^{-1} \sigma | \bar v,\vec{n} \rangle \cr 
&& = {  d_u \over m! |H| } \sum_{ \tau , \sigma } \sum_{ \gamma \in H } 
    \delta ( \tau^{-1} \sigma \gamma )\delta_{\vec{m}\vec{n}}   \Gamma^{(u)}_{  j k  } ( \tau ) 
 B^{u\to 1_H}_{k\nu}  \Gamma^{(s)}_{ i l }    ( \sigma  ) 
 B^{s\to 1_H}_{l\mu }
\eea
The $ | v_i > $ in $V_p$ are normalized as $ < v_i | v_j  > = \delta_{ij} $, so that the $| \bar v , \vec m > $ defined in (\ref{defvbar}) obey   
\bea 
< \bar v,\vec{m} |\sigma |  \bar v,\vec{n} > = \delta_{\vec{m},\vec{n}}\sum_{ \gamma \in H } \delta (  \sigma \gamma ) 
\eea
A permutation outside $ H$  would lead to overlaps $< v_i |v_j >$ for $ i \ne j $, which is zero. 
Thus, we have 
\bea 
\langle  \vec{m}, u,  \nu ; j| \vec{n} , s ,  \mu ; i \rangle 
&&  =  { 1 \over m! |H| } d_u \sum_{ \tau , \sigma }  \sum_{ \gamma \in H } 
 \Gamma^{(u)}_{  j k  } ( \sigma \gamma  ) 
 B^{u\to 1_H}_{k\nu}  \Gamma^{(s)}_{ i l }    ( \sigma  ) 
 B^{s\to 1_H}_{l \mu } \delta_{\vec{m}\vec{n}} \cr 
&& =  {1 \over m! |H| } d_u \sum_{  \sigma }  \sum_{ \gamma \in H } 
 \Gamma^{(u)}_{  j j_1  } ( \sigma )   \Gamma^{(u)}_{  j_1 k  }( \gamma  ) 
 B^{u \to 1_H}_{k\nu }  \Gamma^{(s)}_{ i l }    ( \sigma  ) 
 B^{s\to 1_H}_{l \mu }\delta_{\vec{m}\vec{n}} \cr 
&& =   {1 \over |H| }  \delta_{su} \delta_{j i  } \delta_{ j_1 l  }  \sum_{ \gamma \in H }   
   \Gamma^{(u)}_{  j_1 k  }( \gamma  ) B^{u\to 1_H}_{k\nu }  B^{s\to 1_H}_{l \mu} \delta_{\vec{m}\vec{n}}\cr 
&& =     {1 \over |H| }  \delta_{ ij } \delta_{ su }  \sum_{ \gamma \in H }   
   \Gamma^{(u)}_{  l  k  }( \gamma  ) B^{u\to 1_H}_{k\nu }  B^{s\to 1_H}_{l\mu } \delta_{\vec{m}\vec{n}}\cr 
&& =  \delta_{ ij }\delta_{ su  } B^{u\to 1_H}_{l\alpha } B^{u\to 1_H}_{k\alpha} B^{u\to 1_H}_{k\nu } B^{u\to 1_H}_{l \mu}\delta_{\vec{m}\vec{n}} \cr 
&& = \delta_{ ij }\delta_{ su } \delta_{ \mu \nu } \delta_{\vec{m}\vec{n}}
\eea 
which completes the demonstration.

We will now calculate the matrix elements of $D$ in the Gauss graph basis, showing how  the matrix $M^{(ij)}_{s\mu_1\mu_2 ; u\nu_1\nu_2}$ 
appears. 
\bea
\langle O_{T,t}^\dagger (\sigma_2)DO_{R,r}(\sigma_1) \rangle &=&
{|H|^2\over m!}\sum_{s,u\vdash m}\sum_{\mu_1\mu_2\nu_1\nu_2}\sqrt{d_s d_u}
   \Gamma^{(s)}_{jk}(\sigma_2) B^{s\to 1_H}_{j\mu_1}B^{s\to 1_H}_{k\mu_2}
   \Gamma^{(u)}_{lm}(\sigma_1) B^{u\to 1_H}_{l\nu_1}B^{u\to 1_H}_{m\nu_2}\cr
   &&\qquad\times
   \langle O^\dagger_{T,(t,u)\nu_1\nu_2}D O_{R,(r,s)\mu_1\mu_2}\rangle\cr
   &=&-{|H|^2\over m!}\sum_{s,u\vdash m}\sum_{\mu_1\mu_2\nu_1\nu_2}
\sqrt{d_s d_u}
   \Gamma^{(s)}_{jk}(\sigma_2) B^{s\to 1_H}_{j\mu_1}B^{s\to 1_H}_{k\mu_2}
   \Gamma^{(u)}_{lm}(\sigma_1) B^{u\to 1_H}_{l\nu_1}B^{u\to 1_H}_{m\nu_2}\cr
   &&\!\!\!\!\times
   \sum_{i<j}g_{YM}^2 M^{(ij)}_{s\mu_1\mu_2 ; u \nu_1\nu_2}
\Delta^{R,r ;  T,t}_{ij}  \cr
   &=&-{|H|^2\over m!}g_{YM}^2\sum_{s,u\vdash m}\sum_{\mu_1\mu_2\nu_1\nu_2}
   \Gamma^{(s)}_{jk}(\sigma_2) B^{s\to 1_H}_{j\mu_1}B^{s\to 1_H}_{k\mu_2}
   \Gamma^{(u)}_{lm}(\sigma_1) B^{u\to 1_H}_{l\nu_1}B^{u\to 1_H}_{m\nu_2}\cr
   &&\!\!\!\!\times
   \sum_{i<j} \Delta^{R,r ;  T,t}_{ij} m\left(
       \langle \vec{m},s,\mu_2\, ;\, a|E^{(1)}_{ii}|\vec{m},u,\nu_2\, ;\, b\rangle
       \langle \vec{m},u,\nu_1\, ;\, b|E^{(1)}_{jj}|\vec{m},s,\mu_1\, ;\, a\rangle\right.\cr
  &&\!\!\!\!\left. +
       \langle \vec{m},s  ,\mu_2\, ;\, a|E^{(1)}_{jj}|\vec{m},u,\nu_2\, ;\, b\rangle
       \langle \vec{m},u,\nu_1\, ;\, b|E^{(1)}_{ii}|\vec{m},s,\mu_1\, ;\, a\rangle\right)
 \label{bigtocompute}
\eea
Focus on the evaluation of
\bea
&&\sum_{s,u\vdash m}\sum_{\mu_1\mu_2\nu_1\nu_2}
   \Gamma^{(s)}_{jk}(\sigma_2) B^{s\to 1_H}_{j\mu_1}B^{s\to 1_H}_{k\mu_2}
   \Gamma^{(u)}_{lm}(\sigma_1) B^{u\to 1_H}_{l\nu_1}B^{u\to 1_H}_{m\nu_2}\cr
&&m\left(
       \langle \vec{m},s,\mu_2\, ;\, a|E^{(1)}_{ii}|\vec{m},u,\nu_2\, ;\, b\rangle
       \langle \vec{m},u,\nu_1\, ;\, b|E^{(1)}_{jj}|\vec{m},s,\mu_1\, ;\, a\rangle\right.\cr
  &&\qquad\left. +
       \langle \vec{m},s,\mu_2\, ;\, a|E^{(1)}_{jj}|\vec{m},u,\nu_2\, ;\, b\rangle
       \langle \vec{m},u,\nu_1\, ;\, b|E^{(1)}_{ii}|\vec{m},s,\mu_1\, ;\, a\rangle\right)
 \label{tocompute}
\eea
To evaluate (\ref{tocompute}) start by considering
\bea 
&& \sum_u  |\vec{m}, u , \nu_2 ; b \rangle \langle \vec{m},u , \nu_1 ; b | 
B_{l\nu_1}^{u\to 1_H} B_{m\nu_2}^{u\to 1_H} \Gamma_{l m }^{ (u) } 
 ( \sigma_2 ) \cr 
&&  ={ 1 \over |H| m! }   \sum_u \sum_{ \sigma , \tau \in S_m } 
   d_u ~~ B^{u\to 1_H}_{j\nu_2} \Gamma^{(u)}_{bj} (\sigma ) |v_{\sigma} \rangle 
   \langle v_{ \tau } | \Gamma^{(u)}_{bk} (\tau) B^{u\to 1_H}_{k\nu_1} 
B_{l \nu_1 }^{u\to 1_H } B_{m\nu_2 }^{u \to 1_H} \Gamma_{l m }^{ (u) } 
 ( \sigma_2 ) \cr 
&& = { 1 \over |H| m! }  \sum_u \sum_{ \sigma , \tau \in S_m }  d_u ~~ 
 | v_{ \sigma } \rangle \langle v_{ \tau } | 
\Gamma^{(u)}_{jk}(\sigma^{-1} \tau)B^{u\to 1_H}_{j \nu_2} B^{u\to 1_H}_{k \nu_1} B^{u\to 1_H}_{l\nu_1} B^{u\to 1_H}_{m \nu_2} \Gamma_{lm}(\sigma_2)\cr 
&& = { 1 \over |H| m! } \sum_u  \sum_{ \sigma , \tau \in S_m }  { 1 \over |H|^2 }  \sum_{ \gamma_1 , \gamma_2 \in H }  
 d_u \Gamma_{jm}^{(u)} ( \gamma_1 ) \Gamma_{ kl }^{(u)} ( \gamma_2 ) 
\Gamma_{ lm }^{(u)} ( \sigma_2 ) \Gamma_{ jk }^{(u)} ( \sigma^{-1}  \tau ) 
| v_{ \sigma } \rangle \langle v_{ \tau } | \cr 
&& =  { 1 \over m!  | H | } \sum_u \sum_{ \sigma , \tau \in S_m }  { 1 \over |H|^2 } \sum_{ \gamma_1 , \gamma_2 \in H } 
    d_u \chi_u ( \gamma_1 \sigma_2^{-1} \gamma_2^{-1} \tau^{-1} \sigma )     | v_{ \sigma } \rangle \langle v_{ \tau } |  \cr 
&& = { 1 \over |H|^3 }  \sum_{ \sigma , \tau \in S_m } \sum_{ \gamma_1 , \gamma_2 \in H }
\delta (  \gamma_1 \sigma_2^{-1} \gamma_2^{-1} \tau^{-1} \sigma  ) 
| v_{ \sigma } \rangle \langle v_{ \tau } |
\eea 
Using this twice we get for the first term in (\ref{tocompute})
\bea\label{usetwice} 
T_1 &=&  { m \over |H|^6 }
 \sum_{ \gamma_1 \cdots \gamma_4  } \sum_{ \alpha , \beta , \sigma , \tau } 
\langle v_{ \beta } |   E_{ii}^{(1)} |v_{\sigma} \rangle \langle v_{\tau}  | E_{jj}^{(1)} |  v_{\alpha }   \rangle    
 \delta (  \gamma_1 \sigma_2^{-1} \gamma_2^{-1} \tau^{-1} \sigma  ) 
 \delta ( \gamma_3 \sigma_1 \gamma_4^{-1} \beta^{-1} \alpha  ) \cr 
 &=& { m \over |H|^6 } \sum_{\beta , \tau  }
 \sum_{ \gamma_i }  \langle \bar v | \beta^{-1} E_{ii}^{(1)} 
\tau \gamma_2 \sigma_2  \gamma_1^{-1}  
  | \bar v \rangle  ~~ 
\langle \bar v | \tau^{-1} E_{jj}^{ (1) }
\beta \gamma_4 \sigma_1^{-1}\gamma_3^{-1}  
 | \bar v \rangle \cr 
&=&  { m \over |H|^4 }  \sum_{\beta , \tau  } 
\sum_{ \gamma_2, \gamma_4}   
  \langle \bar v |  E_{ii}^{\beta^{-1} (1)} \beta^{-1} 
\tau \gamma_2 \sigma_2
  | \bar v \rangle  ~~ \langle \bar v |  E_{jj}^{ \tau^{-1} (1) }
 \tau^{-1}\beta \gamma_4 \sigma_1^{-1}
 | \bar v \rangle
\eea
We dropped the $\gamma_1 , \gamma_3$ and and picked up $|H|^2$ using invariance of $ | \bar v \rangle $ under $H$. 

Now consider $ E^{ \sigma(1) } | \bar v \rangle $ ( or equivalently 
$ \langle \bar v |  E^{ \sigma^{-1} (1) }$). This gives $ | \bar v \rangle $
if $ \sigma(1)$ belongs to the set $S_i$ of integers
between $ m_1 + m_2 + \cdots m_{i-1} +1 $ and  $ m_1 + m_2 + \cdots + m_{i}$ both inclusive. The above expression 
will be zero unless $ \beta^{-1} ( 1) \in S_i $ and 
$ \tau^{-1} ( i ) \in S_j $. We also note that 
\bea 
\langle \bar v | \sigma | \bar v \rangle  = \sum_{ \gamma \in H } 
\delta ( \sigma \gamma ) 
\eea
So we can write 
\bea 
T_1 & = &  {m\over |H|^4} \sum_{ \beta , \tau } \sum_{ \gamma_i } 
\delta ( \beta^{-1} \tau \gamma_2 \sigma_2 \gamma_3 ) \delta ( \tau^{-1} \beta \gamma_4 \sigma_1^{-1}  \gamma_1 ) \cr 
&& \qquad \sum_{ k  \in S_i } \delta ( \beta^{-1} ( 1 ) , k  ) 
 \sum_{ l   \in S_j } \delta ( \tau^{-1} (1) , l ) 
\eea  
The delta functions in the second line imply 
\bea 
&& \beta^{-1} \tau ( l  ) = k,   \cr 
&&  \tau^{-1} \beta ( k ) = l  ; ~ \hbox{ for }   l \in S_j , k \in S_i 
\eea
We can replace the two delta functions in the last line with 
a delta function constraining $ \beta^{-1} \tau $, i.e 
$ \sum_{ l \in S_j } \sum_{ k \in S_i } \delta ( \beta^{-1} \tau (k) , l )$. 
This can be done in the current context, because the rest of the expression 
only depends on $\beta^{-1} \tau $. 
If we replace  $\beta \rightarrow \beta^{-1} \alpha   ; \tau^{-1} \rightarrow \tau^{-1} \alpha  $ with
 $\alpha \in Z_m $, this amounts to replacing the $1$ by $\alpha (1)$. 
By summing over $\alpha $ in $Z_m $ we can replace the $1$ by a sum over 
$i$ from $1$ to $m$ (normalized by $1/m$). 
So we are lead to write 
\bea 
T_1 &= &  { 1 \over  |H|^4 } \sum_{ \beta , \tau } \sum_{ \gamma_i \in H } 
\delta ( \beta^{-1} \tau \gamma_2 \sigma_2  \gamma_3 )   ~~ 
\delta ( \tau^{-1} \beta \gamma_4 \sigma_1^{-1}  \gamma_1 )
\sum_{ l \in S_j } \sum_{ k \in S_i } \delta ( \beta^{-1} \tau (k) , l ) 
 \cr 
& =& { m!  \over  |H|^4 }  \sum_{ \beta } \sum_{ \gamma_i \in H } 
\delta ( \beta^{-1}  \gamma_2 \sigma_2  \gamma_3 )   ~~ 
\delta (  \beta \gamma_4 \sigma_1^{-1}  \gamma_1 )   
\sum_{ l \in S_j } \sum_{ k \in S_i } \delta ( \beta^{-1} (k) , l )  \cr 
&=&  {  m !  \over |H|^4 }  \sum_{ \beta } \sum_{ \gamma_i \in H } 
\delta ( \beta^{-1}  \gamma_2 \sigma_2  \gamma_3 ) ~~ \delta (  \beta \gamma_4 \sigma_1^{-1}  \gamma_1 ) ~~   
n^+_{ij} ( \beta^{-1} ) \cr 
&=&  {  m!  \over |H|^4 } \sum_{ \gamma_i } \delta ( \gamma_2 \sigma_2 \gamma_3 \gamma_4 \sigma_1^{-1}  \gamma_1 ) ~~ n^+_{ij} ( \gamma_4 \sigma_1^{-1}  \gamma_1 ) \cr 
&=&  { m! \over |H|^2  }  \sum_{ \gamma_1, \gamma_2 } 
\delta ( \gamma_1 \sigma_2 \gamma_2  \sigma_1^{-1}  ) ~~ n^+_{ij} (  \sigma_1  )
\eea
We have recognized the definition of $n^+_{ij} ( \sigma ) $ and 
the fact that it is invariant under left and right multiplication by $H$. 
In the second term of (\ref{tocompute}) we have $i,j$ exchanged and $n^+_{ij } = n^{-}_{ ji} $.
Combining the two terms we would get 
\bea 
 { m! \over |H|^2  }  \sum_{ \gamma_1, \gamma_2 } 
\delta ( \gamma_1 \sigma_2 \gamma_2  \sigma_1^{-1}  ) ~~ n_{ij} (  \sigma_1  )
\eea
Plugging this into (\ref{bigtocompute}) we find
\bea
\langle O_{T,s}^\dagger (\sigma_2)O_{R,r}(\sigma_1) \rangle =
   - g_{YM}^2\sum_{ \gamma_1, \gamma_2 } 
\delta ( \gamma_1 \sigma_2 \gamma_2  \sigma_1^{-1}  )\sum_{i<j}   ~~ n_{ij} (  \sigma_1  )
  \Delta^{R,r ;  T,s}_{ij}  
 \label{bigcomputed}
\eea
which proves that the Gauss graph operators indeed diagonalize the impurity labels.
We can also write (\ref{bigcomputed}) as
\bea
D O_{R,r}(\sigma_1)  =
   - g_{YM}^2 \sum_{i<j}   ~~ n_{ij} (  \sigma_1  )
 \Delta_{ij}  O_{R,r}(\sigma_1)
 \label{lovelyanswer}
\eea
This last eigenproblem has been considered in detail in \cite{gs}. 
Taking a large $N$ continuum limit, the above discrete problem becomes a differential equation, equivalent to a
set of decoupled oscillators. 
The same spectrum is obtained by solving the discrete problem or the large $N$ continuum differential equation.

The discussion above has focused on the case that $R$ has $p$ long rows. 
These operators are dual to giant gravitons wrapping an S$^3\subset$AdS$_5$.
The case that $R$ has $p$ long columns, which is dual to giant gravitons wrapping an S$^3\subset$S$^5$,
is easily obtained from the above results. 
The $\Delta_{ij}$ for this case is obtained by replacing the $r_i\to -r_i$ and $r_j\to -r_j$ in (\ref{0term}), (\ref{pterm})
and (\ref{mterm}).
The final result (\ref{lovelyanswer}) is unchanged when written in terms of the new $\Delta_{ij}$.

\section{ Outlook }

There are a number of natural ways in which this work can be extended.
We have limited ourselves to restricted Schur polynomials labeled by
Young diagrams $R$ that have well separated corners, corresponding to
giant gravitons that are well separated in spacetime. 
We conjecture that the permutation $\sigma $ specifying the brane-string configuration obeying the Gauss Law, and appearing in the operators $\cO_{R,r} ( \sigma ) $will continue to provide a diagonalization of
the  dilatation operator to all orders in the loop expansion
in this distant corners limit. The action will be diagonal in $\sigma $ 
but there will be a mixing of the $R$ label which involves the movement
of  more boxes at higher orders. Proving (or disproving) this conjecture 
would give important information on the structure of higher loop corrections to the dilatation operator.

Another  fascinating  generalization is  to consider is the case where 
some of the branes are coincident, in which case some of the
row lengths of $R$ will be equal.  
This case is particularly interesting as it
corresponds to non-abelian brane worldvolume theories. 
A first step would be to give a general account of the counting of 
restricted Schurs in terms of the Gauss Law for these non-abelian brane 
worldvolumes. For initial studies in this 
direction see \cite{Balasubramanian:2004nb}. 
In line with the counting to construction philosophy we have followed
in this article, a general proof of this counting should contain the 
hints of the corresponding operator construction. Implementing this will 
require some work in making the action  of the one-loop  dilatation 
operator more explicit.

The counting of BPS states in \cite{pasram}
was expressed  in terms of bit strings $Y^k$, 
built using $k$-bits at a time. 
In this article we have focused on a description of strings 
by assembling single bit strings. The precise relation between these 
two descriptions will be interesting to clarify.

We have considered operators built from $Z$s and 
dilutely doped with a single type of impurity $Y$. 
The one loop dilatation operator has same form in the $sl(2)$ sector \cite{deMelloKoch:2011vn}, where we
dope with covariant derivatives, so our
 double coset ansatz works in that case too. 
In general we could build operators with impurities that include more
types of scalars together with covariant derivatives and fermions. 
This would allow a complete description of the one loop, large $N$ but non-planar dilatation operator. 
The one loop planar dilatation operator is integrable\cite{beisertoneloop}.
Is this complete one loop non-planar dilatation operator
integrable in this sector of perturbations around well-separated half-BPS giants? 
 Is there a double coset ansatz that can be used to diagonalize the problem?

An interesting concept we have found very useful 
in this paper is what we may call the  
{\it counting to construction philosophy}. 
This is  the expectation that once we have proved that  
some  framework based on groups (e.g permutation  groups) or algebras (e.g Brauer algebras)  correctly counts the quantum states, 
expected from gauge-string duality for example, then the same framework will contain the information for constructing the states, often via tools related to Fourier transformation on the groups or algebras along, frequently, 
with Schur-Weyl duality. The link between 
enumeration and construction is also an active theme of research 
in areas such as the mathematical classification of molecular structures 
using double cosets, see for example \cite{klw}.
This theme  also appears  in the categorification of numeric 
to homological invariants in the context 
of knot theory \cite{khovanov} and branes \cite{wittenknot}, 
with interesting links to 
Schur-Weyl duality and representation theory \cite{FKS}. It is clear that 
there is much to be understood about the interplay of this theme 
with gauge-string duality.

\section*{Acknowledgements}

We thank Jurgis Pasukonis and Congkao Wen for useful discussions. SR 
is  supported by an STFC Standard Grant ST/J000469/1, ``String theory, gauge theory, and duality.'' 
RdMK is supported by the South African Research Chairs
Initiative of the Department of Science and Technology and National Research Foundation.

\begin{appendix}

\section{Conventions}\label{nutshell}

In this Appendix we will spell out our conventions for $(\chi_{R,(r,s)\mu\nu}(Z,Y))^\dagger $. It is straight forward
to check that
$$
  {\rm Tr}\left(\sigma Z^{\otimes\, n}Y^{\otimes\, m}\right)^\dagger =
   {\rm Tr}\left(\sigma^{-1} Z^{\dagger\,\otimes\, n}Y^{\dagger\,\otimes\, m}\right)
$$
Using this we find
\bea
  \chi_{R,(r,s)\mu\nu}(Z,Y)^\dagger
 &&={1\over n!m!}\sum_{\sigma\in S_{n+m}}{\rm Tr}_{R}\left( P_{R\to (r,s)\mu\nu}\Gamma_R (\sigma)\right)
                        {\rm Tr}_{V_N^{\otimes\, n+m}}\left(\sigma^{-1}\,Z^{\dagger\,\otimes\, n}Y^{\dagger\,\otimes\, m}\right)\cr
 &&={1\over n!m!}\sum_{\sigma\in S_{n+m}}{\rm Tr}_{R}\left( P_{R\to (r,s)\mu\nu}\Gamma_R (\sigma^{-1})\right)
                        {\rm Tr}_{V_N^{\otimes\, n+m}}\left(\sigma\,Z^{\dagger\,\otimes\, n}Y^{\dagger\,\otimes\, m}\right)\cr
 &&={1\over n!m!}\sum_{\sigma\in S_{n+m}}{\rm Tr}_{R}\left( P_{R\to (r,s)\mu\nu}\Gamma_R (\sigma)^T \right)
                        {\rm Tr}_{V_N^{\otimes\, n+m}}\left(\sigma\,Z^{\dagger\,\otimes\, n}Y^{\dagger\,\otimes\, m}\right)\cr
 &&={1\over n!m!}\sum_{\sigma\in S_{n+m}}{\rm Tr}_{R}\left( P_{R\to (r,s)\mu\nu}^T\Gamma_R (\sigma)\right)
                        {\rm Tr}_{V_N^{\otimes\, n+m}}\left(\sigma\,Z^{\dagger\,\otimes\, n}Y^{\dagger\,\otimes\, m}\right)\cr
 &&={1\over n!m!}\sum_{\sigma\in S_{n+m}}{\rm Tr}_{R}\left( P_{R\to (r,s)\nu\mu}\Gamma_R (\sigma)\right)
                        {\rm Tr}_{V_N^{\otimes\, n+m}}\left(\sigma\,Z^{\dagger\,\otimes\, n}Y^{\dagger\,\otimes\, m}\right)
\eea
Thus, following the original derivation of the two point function \cite{Bhattacharyya:2008rb} we find
\bea
\langle \chi_{T,(t,u)\alpha\beta}\chi_{R,(r,s)\mu\nu}(Z,Y)^\dagger \rangle &&= n!m!{\rm Tr}(P_{T\to (t,u)\alpha\beta}P_{R\to (r,s)\nu\mu})\cr
&&={{\rm hooks}_R\over {\rm hooks}_t {\rm hooks}_u}\delta_{TR}\delta_{rt}\delta_{us}\delta_{\beta\nu}\delta_{\alpha\mu}
\eea
This is the convention followed in this article, and it matches \cite{Bhattacharyya:2008rb}. Our motivation for adopting this 
convention, is that we get the
natural orthogonality (\ref{cosettwopoint}) between Gauss graph operators. In \cite{Koch:2011hb} a different definition 
\bea
  \chi_{R,(r,s)\mu\nu}(Z,Y)^\dagger\equiv
 {1\over n!m!}\sum_{\sigma\in S_{n+m}}{\rm Tr}_{R}\left( P_{R\to (r,s)\mu\nu}\Gamma_R (\sigma)\right)
                        {\rm Tr}_{V_N^{\otimes\, n+m}}\left(\sigma\,Z^{\dagger\,\otimes\, n}Y^{\dagger\,\otimes\, m}\right)
\eea
was used. This implies
\bea
\langle \chi_{T,(t,u)\alpha\beta}\chi_{R,(r,s)\mu\nu}(Z,Y)^\dagger \rangle 
={{\rm hooks}_R\over {\rm hooks}_t {\rm hooks}_u}\delta_{TR}\delta_{rt}\delta_{us}\delta_{\beta\mu}\delta_{\alpha\nu}
\eea
This convention looks natural if one interprets multiplicity labels as Chan-Paton factors and is the motivation for adopting
this convention in \cite{Koch:2011hb}.

\section{Counting operators}

In this appendix we will give some examples of the counting arguments constructed in section 2.

First we deal with the Gauss graph counting problem. To approach this numerically we have found it
easiest to implement (\ref{ingap}) in GAP. We have counted the number $N_C$ of open string
configurations for the stated $\vec{m}$ shown below.

{\vskip 1.0cm}

\begin{tabular}{|c|c|c|}
\hline
\textbf{Total Number of Strings $(m)$} & \textbf{Valencies $(\vec{m}=\{m_i\})$} & \textbf{Configurations $N_C$}\\
\hline
4 & \{2,1,1\} & 7\\
\hline
5 & \{3,1,1\} & 7\\
\hline
5 & \{4,1\} & 2\\
\hline
5 & \{3,2\} & 3\\
\hline
5 & \{2,2,1\} & 11\\
\hline
3 & \{1,1,1\} & 6\\
\hline
8 &\{4,2,1,1\} & 68\\
\hline
\end{tabular}

{\vskip 1.0cm}

To count the number of restricted Schur polynomials, according to (\ref{countrest}) we should sum the squares of 
the Kostka numbers. The Kostka numbers are easily evaluated with the help, for example, of the Symmetrica
program\cite{symmetrica}. We will write the Kostka numbers as
$$
  {\tiny \yng(2)}\otimes {\tiny \yng(1)}\otimes {\tiny \yng(1)} ={\tiny \yng(4)} \, \oplus \, 2{\tiny \yng(3,1)} \, \oplus 
                          \, {\tiny \yng(2,1,1)} \, \oplus \, {\tiny \yng(2,2)}
$$
The left hand side of this equation determines $\vec{m}=(2,1,1)$. The right hand side shows the non zero irreps $s$ that we can
obtain. The coefficient of each term is the Kostka number. Thus, for example ${\cal K}_{{\tiny \yng(3,1)},(2,1,1)}=2$.
For each line of the table above it is now a simple matter to check that we reproduce $N_C$:
$$
  {\tiny \yng(2)}\otimes {\tiny \yng(1)}\otimes {\tiny \yng(1)} ={\tiny \yng(4)} \, \oplus \, 2{\tiny \yng(3,1)} \, \oplus 
                          \, {\tiny \yng(2,1,1)} \, \oplus \, {\tiny \yng(2,2)}
$$
$$
  1^2 + 2^2 + 1^2 + 1^2 = 7
$$

$$
  {\tiny \yng(3)}\otimes {\tiny \yng(1)}\otimes {\tiny \yng(1)} ={\tiny \yng(5)} \, \oplus \, 2{\tiny \yng(4,1)} \, \oplus 
                          \, {\tiny \yng(3,1,1)} \, \oplus \, {\tiny \yng(3,2)}
$$
$$
  1^2 + 2^2 + 1^2 + 1^2 = 7
$$
$$
  {\tiny \yng(4)}\otimes {\tiny \yng(1)} ={\tiny \yng(5)} \, \oplus \, {\tiny \yng(4,1)}
$$
$$
  1^2 + 1^2  = 2
$$

$$
  {\tiny \yng(3)}\otimes {\tiny \yng(2)} ={\tiny \yng(5)} \, \oplus \, {\tiny \yng(4,1)} \, \oplus \, {\tiny \yng(3,2)}
$$
$$
  1^2 + 1^2 + 1^2 = 3
$$

$$
  {\tiny \yng(2)}\otimes {\tiny \yng(2)}\otimes {\tiny \yng(1)} ={\tiny \yng(5)} \, \oplus \, 2{\tiny \yng(4,1)} \, \oplus 
                          \, {\tiny \yng(3,1,1)} \, \oplus \, 2 {\tiny \yng(3,2)} \, \oplus\, {\tiny \yng(2,2,1)}
$$
$$
  1^2 + 2^2 + 1^2 + 2^2  + 1^2 = 11
$$

$$
  {\tiny \yng(1)}\otimes {\tiny \yng(1)}\otimes {\tiny \yng(1)} ={\tiny \yng(3)} \, \oplus \, 2{\tiny \yng(2,1)} \, \oplus 
                          \, {\tiny \yng(1,1,1)} 
$$
$$
  1^2 + 2^2 + 1^2 = 6
$$

$$
  {\tiny \yng(4)}\otimes {\tiny \yng(2)}\otimes {\tiny \yng(1)}\otimes {\tiny \yng(1)} ={\tiny \yng(8)} \, \oplus \, 3{\tiny \yng(7,1)} \, \oplus 
                          \, 3 {\tiny \yng(6,1,1)} \, \oplus 
                          \, 4 {\tiny \yng(6,2)}  \, \oplus 
                          \, 3 {\tiny \yng(5,3)}  \, \oplus 
                          \, 4 {\tiny \yng(5,2,1)}
$$
$$
                          \, \oplus 
                          \,  {\tiny \yng(5,1,1,1)}\, \oplus 
                          \,  {\tiny \yng(4,4)}\, \oplus 
                          \, 2 {\tiny \yng(4,3,1)}\, \oplus 
                          \, {\tiny \yng(4,2,1,1)}\, \oplus 
                          \, {\tiny \yng(4,2,2)}
$$
$$
1^2 + 3^2 + 3^2 + 4^2 + 3^2 + 4^2 + 1^2 + 1^1 + 2^2 + 1^2 + 1^2 = 68
$$

\section{Examples of the Gauss graph operators}

In this section we will use (\ref{ggo}) to explicitly construct some examples of Gauss operators.
We have two goals in mind: to demonstrate how the formula (\ref{ggo}) is used and to make contact 
with operators already constructed in the literature.

\subsection{BPS Operators}

The BPS operator is associated with the open string configuration that has all strings looping back 
to the brane they start from. 
This corresponds to taking the identity for $\sigma$. 
In this case
$$
  \Gamma^{(s)}_{jk}({\bf 1})=\delta_{jk}
$$
so that
$$
  O({\bf 1})={|H|\over m!}\sum_{s\vdash m}\sum_{\mu_1,\mu_2}\sqrt{d_s}B^{s \rightarrow 1_H }_{j \mu_1 } B^{s \rightarrow 1_H }_{j \mu_2 } O_{R,(r,s)\mu_1\mu_2}
            ={|H|\over m!}\sum_{s\vdash m}\sum_{\mu}\sqrt{d_s} O_{R,(r,s)\mu\mu}
$$
This is exactly what \cite{DeComarmond:2010ie} has found based on numerical studies.

\subsection{Two row operators}

In this case we have no multiplicity label so that, up to a normalization factor of ${|H|\over m!}$ which we drop, 
we have
\bea
  O(\sigma)&=&\sum_{s\vdash m}\sqrt{d_s}\Gamma^{(s)}_{jk}(\sigma )B^{s \rightarrow 1_H }_{j}B^{s \rightarrow 1_H }_{k} O_{R,(r,s)}\cr
           &=&{1\over\prod_i m_i!}\sum_{s\vdash m}
            \sum_{\alpha\in \prod_i S_{m_i}}\sqrt{d_s}
            \Gamma^{(s)}_{jk}(\sigma )\Gamma^{(s)}_{kj}(\alpha) O_{R,(r,s)}\cr
  &=&{1\over\prod_i m_i!}\sum_{s\vdash m}\sum_{\alpha\in \prod_i S_{m_i}}\sqrt{d_s}\chi_{s}(\sigma \alpha) O_{R,(r,s)}
\eea
where $\chi_{s}(\sigma )$ is the character of $\sigma\in S_m$ in irrep $s$. 

Consider $m=3$, $m_1=1$ and $m_2=2$. 
There are two possible Gauss operators.
For $\sigma=1$ we obtain the BPS operator as discussed above.
The other configuration, which is non-BPS, is obtained for $\sigma =(23)$. In this case
$$
  {1\over 2!}{\rm Tr}\left( (\Gamma^{({\tiny\yng(3)})}(1)+ \Gamma^{({\tiny\yng(3)})}((12))\Gamma^{({\tiny\yng(3)})}((23))\,\right)=1
$$
$$
  {1\over 2!}{\rm Tr}\left( (\Gamma^{({\tiny\yng(2,1)})}(1)+\Gamma^{({\tiny\yng(2,1)})}((12))\Gamma^{({\tiny\yng(2,1)})}((23))\,\right)=-{1\over 2}
$$
so that
$$
  O((23))=1\cdot O_{R,(r,{\tiny\yng(3)})}+\left(-{1\over 2}\right)\cdot\sqrt{2}O_{R,(r,{\tiny \yng(2,1)})}
         = O_{R,(r,{\tiny\yng(3)})}-{1\over \sqrt{2}}O_{R,(r,{\tiny \yng(2,1)})}
$$
which is in perfect agreement with section 5.1 of \cite{DeComarmond:2010ie}. 
Now consider $m=4$ and $m_1=2$, $m_2=2$. 
There are three possible Gauss operators.
For $\sigma = {\bf 1}$ we again obtain the BPS operator. For $\sigma =(23)$, using 
$$
  {1\over 2! 2!}{\rm Tr}\left( (\Gamma^{({\tiny\yng(4)})}(1)+\Gamma^{({\tiny\yng(4)})}((12))
                               +\Gamma^{({\tiny\yng(4)})}((34))+\Gamma^{({\tiny\yng(4)})}((12)(34)))
             \Gamma^{({\tiny\yng(4)})}((23))\,\right)=1
$$
$$
  {1\over 2! 2!}{\rm Tr}\left((\Gamma^{({\tiny\yng(3,1)})}(1)+\Gamma^{({\tiny\yng(3,1)})}((12))
                              +\Gamma^{({\tiny\yng(3,1)})}((34))+\Gamma^{({\tiny\yng(3,1)})}((12)(34)))
                   \Gamma^{({\tiny\yng(3,1)})}((23))\,\right)=0
$$
$$
  {1\over 2! 2!}{\rm Tr}\left((\Gamma^{({\tiny\yng(2,2)})}(1)+\Gamma^{({\tiny\yng(2,2)})}((12))+\Gamma^{({\tiny\yng(2,2)})}((34))
                              +\Gamma^{({\tiny\yng(2,2)})}((12)(34)))
               \Gamma^{({\tiny\yng(2,2)})}((23))\,\right)=-{1\over 2}
$$
we find
$$
  O((23)) = O_{R,(r,{\tiny\yng(4)})}-{1\over \sqrt{2}}O_{R,(r,{\tiny \yng(2,2)})}
$$
The last configuration is obtained for $\sigma =(14)(23)$. In this case
$$
  {1\over 2! 2!}{\rm Tr}\left((\Gamma^{({\tiny\yng(4)})}(1)+\Gamma^{({\tiny\yng(4)})}((12))+\Gamma^{({\tiny\yng(4)})}((34))
                              +\Gamma^{({\tiny\yng(4)})}((12)(34)))\Gamma^{({\tiny\yng(4)})}((14)(23))\,\right)=1
$$
$$
  {1\over 2! 2!}{\rm Tr}\left((\Gamma^{({\tiny\yng(3,1)})}(1)+\Gamma^{({\tiny\yng(3,1)})}((12))
                              +\Gamma^{({\tiny\yng(3,1)})}((34))+\Gamma^{({\tiny\yng(3,1)})}((12)(34)))
                \Gamma^{({\tiny\yng(3,1)})}((14)(23))\,\right)=-1
$$
$$
  {1\over 2! 2!}{\rm Tr}\left((\Gamma^{({\tiny\yng(2,2)})}(1)+\Gamma^{({\tiny\yng(2,2)})}((12))
                              +\Gamma^{({\tiny\yng(2,2)})}((34))+\Gamma^{({\tiny\yng(2,2)})}((12)(34)))
          \Gamma^{({\tiny\yng(2,2)})}((14)(23))\,\right)=1
$$
so that
$$
  O((23)) = O_{R,(r,{\tiny\yng(4)})}+\sqrt{2} O_{R,(r,{\tiny \yng(2,2)})}-\sqrt{3}O_{R,(r,{\tiny\yng(3,1)})}
$$
These results are in complete agreement with section 5.2 of \cite{DeComarmond:2010ie}.

Next consider removing $m=8$ boxes with $m_1=m_2=4$. 
The relevant dilatation operator equation, given in equation (4.3) of \cite{Carlson:2011hy}, is
\begin{eqnarray}\label{recursion_j3}
 && DO_{j,j^3}(b_{0},b_{1})=g_{YM}^2 \left[-{1\over 2}\left( m-{(m+2)(j^3)^2\over j(j+1)}\right)
\Delta O_{j,j^3}(b_{0},b_{1})\right.\nonumber \\
&& + \sqrt{(m + 2j + 4)(m - 2j)\over (2j + 1)(2j+3)} {(j+j^3 +
1)(j-j^3 + 1) \over 2(j + 1)}
 \Delta O_{j+1,j^3}(b_{0},b_{1})
\nonumber \\
&& \left. +\sqrt{(m + 2j + 2)(m - 2j +2)\over (2j + 1)(2j-1)}
{(j+j^3 )(j-j^3 ) \over 2 j} \Delta O_{j-1,j^3}(b_{0},b_{1})
\right]
\end{eqnarray}
where
\begin{eqnarray}
\Delta O(b_{0},b_{1}) &&=\sqrt{(N+b_{0})(N+b_{0}+b_{1})}(O(b_{0}+1,b_{1}-2)+O(b_{0}-1,b_{1}+2))\nonumber \\
 &&-(2N+2b_{0}+b_{1})O(b_{0},b_{1}).\label{}
\end{eqnarray}
The case we study has operators with $j^3=0$ and $j=0,1,2,3,4$ in the notation of \cite{Carlson:2011hy}. 
The above operator is easily diagonalized numerically giving
eigenvalues $0,2,4,6,8$. It is simple to test that our Gauss graph operators
$$
  O(\sigma )= \sum_{s\vdash 8}\sum_{\alpha\in S_4\times S_4}\sqrt{d_s}\chi_s(\alpha\sigma)O_{R,(r,s)}
$$
are eigenfunctions with the following eigenvalues
\bea
    \sigma = {\bf 1}            \leftrightarrow 0\cr
    \sigma = (45)               \leftrightarrow 2\cr
    \sigma = (45)(63)           \leftrightarrow 4\cr
    \sigma = (45)(63)(72)       \leftrightarrow 6\cr
    \sigma = (45)(63)(72)(81)   \leftrightarrow 8
\eea

\subsection{Three row operators}

\begin{figure}[h]
\begin{center}
\resizebox{!}{5.0cm}{\includegraphics{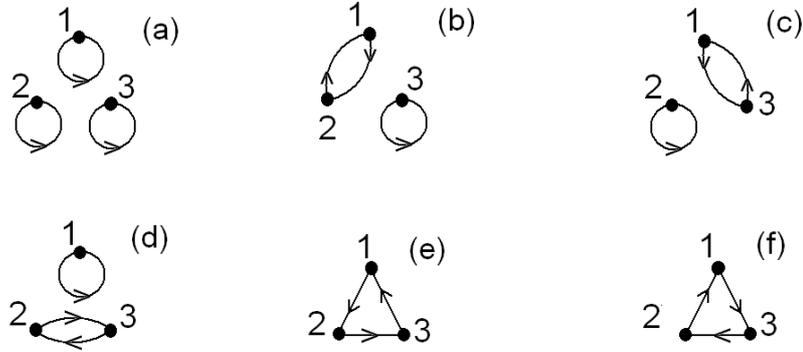}}
\caption{Gauss graphs for three strings and three branes.}
 \label{fig:gauss2}
\end{center}
\end{figure}

Now consider a three row example for which we remove three boxes $m=3$ and $m_1=m_2=m_3=1$. In this case
our Gauss graph operators are
$$
  O(\sigma)=\sum_{s\vdash 3}\sum_{\mu_1,\mu_2}\sqrt{d_s}\Gamma^{(s)}_{jk}(\sigma )
B^{s \to 1_H }_{j \mu_1 }B^{s \to 1_H }_{k \mu_2 } O_{R,(r,s)\mu_1\mu_2}
$$
The subgroup $H=S_1\times S_1\times S_1$ has a single element, the identity. The branching coefficients are thus
$$
  B^{{\tiny \yng(3)} }=1\qquad B^{{\tiny \yng(1,1,1)}}=1
$$
$$
  B^{{\tiny \yng(2,1)}}_{1} =
\left[\begin{matrix} 1\cr 0\end{matrix}\right]\qquad
  B^{{\tiny \yng(2,1)}}_{ 2}=\left[\begin{matrix} 0\cr 1\end{matrix}\right]
$$
Note that we have simplified the notation by dropping from the superscript
the specification $ s \rightarrow  1_H$, 
taking it as understood that these are reduction coefficients 
of the trivial irrep of $H$ appearing in the $s$ 
specified by the Young diagram. 
Each $\sigma\in S_3$ gives a different Gauss graph operator. There are six possible open string configurations
that are given in figure \ref{fig:gauss2} and they correspond as:
\bea
    \sigma = {\bf 1}     \leftrightarrow (a)\cr
    \sigma = (12)         \leftrightarrow (b)\cr
    \sigma = (13)          \leftrightarrow (c)\cr
    \sigma = (23)           \leftrightarrow (d)\cr
    \sigma = (123),(321)     \leftrightarrow (e),(f)
\eea
The last line corresponds to a degeneracy because there are two configurations differing only in orientation
of the open strings. The above correspondence is from comparing to the numerical results of \cite{Koch:2011hb} and 
there is again a perfect agreement. In particular, one can easily test that the above 
Gauss operators are simultaneous eigenfunctions of the matrices $M^{(12)}$, $M^{(23)}$, $M^{(13)}$ of section 3.2 
of \cite{Koch:2011hb}.

\subsection{Four row operators}

Four rows with $m=4$ and $m_1=1=m_2=m_3=m_4$ is also easy to compare. 
The group $H=S_1\times S_1\times S_1\times S_1$ again just has a single element (the identity) so that branching coefficients 
are again trivial to compute
$$
  B^{({\tiny \yng(3)})}=1\qquad B^{({\tiny \yng(1,1,1)})}=1
$$
$$
  B^{{\tiny \yng(2,1)}}_{ 1}=\left[\begin{matrix} 1\cr 0\end{matrix}\right]\qquad
  B^{{\tiny \yng(2,1)}}_{ 2}=\left[\begin{matrix} 0\cr 1\end{matrix}\right]
$$
$$
  B^{{\tiny \yng(3,1)}}_{ 1}=\left[\begin{matrix} 1\cr 0\cr 0\end{matrix}\right]\qquad
  B^{{\tiny \yng(3,1)}}_{ 2}=\left[\begin{matrix} 0\cr 1\cr 0\end{matrix}\right]\qquad
  B^{{\tiny \yng(3,1)}}_{ 3}=\left[\begin{matrix} 0\cr 0\cr 1\end{matrix}\right]
$$
$$
  B^{{\tiny \yng(2,1,1)}}_{ 1}=\left[\begin{matrix} 1\cr 0\cr 0\end{matrix}\right]\qquad
  B^{{\tiny \yng(2,1,1)}}_{  2}=\left[\begin{matrix} 0\cr 1\cr 0\end{matrix}\right]\qquad
  B^{{\tiny \yng(2,1,1)}}_{  3}=\left[\begin{matrix} 0\cr 0\cr 1\end{matrix}\right]
$$
There are 24 operators in this case and here too we have exact, complete, agreement with the
numerical results of \cite{Koch:2011hb}.

\end{appendix}

\end{document}